\documentclass[aps,floatfix,twocolumn,prl]{revtex4}
\usepackage{graphicx,bm,color}
\usepackage{array}
\usepackage{amsmath,amsfonts,amssymb,stackrel}
\usepackage{pdflscape}
\usepackage{hyperref,natbib}
\usepackage{url}
\hypersetup{colorlinks=true}
\hypersetup{urlcolor=blue}
\hypersetup{citecolor=black}
\hypersetup{menucolor=black}
\hypersetup{linkcolor=black}
\usepackage{cleveref}
\usepackage{longtable}
\usepackage{float}
\usepackage{pdfpages}
\usepackage[T1]{fontenc}
\usepackage[utf8]{inputenc}
\usepackage{ulem}

\newcommand{\be}{\begin{equation}}
\newcommand{\ee}{\end{equation}}
\newcommand{\bea}{\begin{eqnarray}}
\newcommand{\eea}{\end{eqnarray}}

\newcommand{\bb}[1]{\left( #1 \right)}

\newcommand{\bbcro}[1]{\left[ #1 \right]}

\newcommand{\ii}{\textrm{i}}
\newcommand{\dd}{\textrm{d}}
\newcommand{\eee}{\textrm{e}}

\newcommand{\qq}{\textbf{q}}
\newcommand{\kk}{\textbf{k}}

\DeclareMathOperator\asin{asin}

\DeclareMathOperator\ch{ch}
\DeclareMathOperator\sh{sh}

\DeclareMathOperator\im{Im}

\begin{document}
\title{{Pair-breaking} collective branch in BCS superconductors and superfluid Fermi gases}

\author{H. Kurkjian, S. N. Klimin and J. Tempere}
\affiliation{TQC, Universiteit Antwerpen, Universiteitsplein 1, B-2610 Antwerpen, Belgi\"e}

\author{Y. Castin}
\affiliation{Laboratoire Kastler Brossel,
École Normale Supérieure, Université PSL, CNRS, Sorbonne Université, Collège de France, Paris, France}

\begin{abstract}
We demonstrate the existence of a collective excitation branch in the pair-breaking continuum of superfluid Fermi gases and BCS superconductors. {At zero temperature,} we analytically continue the equation on the collective mode energy in Anderson's RPA or Gaussian fluctuations through its branch cut associated with the continuum, and obtain the full complex dispersion relation, including in the strong coupling regime. 
{The branch exists as long as the chemical potential $\mu$ is positive and the wavenumber below $\sqrt{2m\mu}/\hbar$ (with $m$ the fermion mass).}
In the long wavelength limit, the branch varies quadratically with the wavenumber, with a complex effective mass that we compute analytically {for an arbitrary interaction strength}.
\end{abstract}
\maketitle

\textit{Introduction --}
Systems with a macroscopic coherence between pairs of fermions exhibit in their excitation spectrum a pair-breaking continuum, whose energy is greater than twice the order-parameter $\Delta$. This is particularly the case of superconductors and cold gases of spin-$1/2$ fermionic atoms. The collective behavior of these systems at energies below $ 2 \Delta $ is known: it is characterized by a bosonic excitation branch, of phononic start in neutral gases \cite{Anderson1958}. The dispersion relation of this branch was calculated \cite{Strinati1998, CKS2006}
and its existence experimentally confirmed \cite{Tachiki1997,Thomas2007,Vale2017}.

Conversely, the existence of a collective mode \textit{inside} the pair-breaking continuum remains a debated question that attracts much interest because of an analogy often suggested with Higgs modes in field theory \cite{Varma2015}. 
{The challenge is to understand {whether} the response of the continuum to an excitation is flat in frequency or presents a nontrivial structure like a resonance.}
We identify two major shortcomings in the existing theoretical treatment {\cite{Schmid1968,Orbach1981,Popov1976,Popov1987-III13,Varma1982}}: (i) it neglects the coupling between the amplitude and phase of the order-parameter, which restricts it to the weak coupling regime, (ii) it is limited to long wavelengths. These shortcomings are prejudicial as they maintain doubts about the very existence of this second collective mode {\cite{Benfatto2015}}, notably at zero wavevector \cite{Stringari2012}.

Here, we clarify the description of the {pair-breaking} collective modes. By analytically continuing the {pair propagator, we reveal a pole below the branch cut associated to the continuum}, for positive chemical potential $\mu> 0$ and nonzero wavenumber only. We obtain the full dispersion relation of this mode {completely accounting for amplitude-phase coupling}. This allows us to deal with the strong coupling regime. {Remarkably, the real part of the branch is wholly below $2 \Delta$ when $ \Delta> 1.210 \mu $ (yet the branch remains separated from the band gap $[0,2\Delta]$ on the real axis by a branch cut)}. In the weak coupling and long wavelength limit, {we agree with the result of \cite{Popov1976} but disagree sharply with the prediction commonly accepted in the literature \cite{Varma1982}}, notably for the damping rate that, we find, has a quadratic start at low wavenumber, {rather than a linear one}. All our predictions are based on {Anderson's RPA or Gaussian approximation} for contact interactions. This theory describes qualitatively well both cold Fermi gases in the BEC-BCS crossover and BCS superconductors ({Coulomb interaction has no effect on amplitude modes at frequencies $O(\Delta/\hbar)$ \cite{Popov1976}}),
and is a prerequisite for any more realistic description of interactions.

The branch we find describes the collective behavior of the pairs following an excitation of their internal degrees of freedom; its frequency is thus not simply the continuum threshold $ 2 \Delta/\hbar $, as for the ``Higgs oscillations'' {predicted and observed} \cite{Kogan1973, Altshuler2006, Gurarie2009, Stringari2012, Foster2015, Klein1980, Shimano2013, Sacuto2014, Devereaux2015, Koehl2018} at zero wavevector.
{It is observable in superfluid Fermi gases as a broadened peak at energies above $2\Delta$ in the order-parameter-amplitude response function.}

\textit{Fluctuations of the order-parameter --}
We consider a homogeneous system of spin-$1/2$ fermions of mass $m$ and chemical potential $\mu$, with contact interactions.
At zero temperature, the fluctuations of the order-parameter $\Delta$ around its equilibrium value admit eigenmodes: the collective modes of the system. Expanding to second order in amplitude $\delta \lambda$ and phase $\delta \theta$ fluctuations yields the Gaussian action \cite{Randeria1997,Randeria2008}
\begin{equation}
	\mathcal{S} = \mathcal{S}_0 + \int \dd \omega \int \dd^3 q 
		\begin{pmatrix}
			 -\ii\Delta\delta\theta^* & \delta\lambda^*
		\end{pmatrix} {{M}}(\omega, \bf{q})
		\begin{pmatrix}
			\ii\Delta\delta\theta \\ \delta\lambda 
		\end{pmatrix}
\end{equation}
The {symmetric} fluctuation matrix $M$ gives access to the propagator of $\Delta$ through a mere inversion. The equation {on} the collective mode energy $z_\qq $ {with} wavevector $\qq$ is then
\be
\mbox{det} M(z_\qq,\qq)=0 
\label{eq:RPA}
\ee
Since the order-parameter $\Delta$ describes pair condensation, the coefficients of its fluctuation matrix contain an integral over the internal wavevector $\kk$ of the pairs, involving $\xi_{\kk} = \hbar ^ 2 k^2 / 2m- \mu $ and $ E_\kk = \sqrt {\xi_\kk^2 + \Delta^2} $, the dispersion relations of free fermions and BCS quasiparticles respectively, as well as the energy $E_{\kk \qq} = E_{\kk+\qq/2} + E_{\kk- \qq/2}$ of a pair of quasiparticles of total wavevector $\qq$:
\bea
\!\! M_{\pm\pm}(z,\qq) \!\! &=& \!\! \int \frac{\dd^3k}{2}\bbcro{\frac{(W_{\kk\qq}^\pm)^2}{z-E_{\kk\qq}}-\frac{(W_{\kk\qq}^\pm)^2}{z+E_{\kk\qq}} +\frac{1}{E_\kk}} \label{eq:Mpp}\\
\!\! M_{+-}(z,\qq)        \!\! &=& \!\! \int \frac{\dd^3k}{2}W_{\kk\qq}^+ W_{\kk\qq}^- \bbcro{\frac{1}{z-E_{\kk\qq}}+\frac{1}{z+E_{\kk\qq}} } \label{eq:Mpm}
\eea
where the indices $+$ and $-$ refer to {phase and amplitude fluctuations} and we introduce the notation $(W_{\kk\qq}^\pm)^2=(E_{\kk+\qq/2}E_{\kk-\qq/2}+\xi_{\kk+\qq/2}\xi_{\kk-\qq/2}\pm\Delta^2)/(2E_{\kk+\qq/2}E_{\kk-\qq/2})$ \footnote{{Here,} $W_{\kk\qq}^+>0$ for all $\kk$ and $W_{\kk\qq}^->0$ {iff} $k^2>2m\mu/\hbar^2-q^2/4$.}. {Eqs.~(\ref{eq:RPA}--\ref{eq:Mpm}) are} found also with RPA \cite{Anderson1958,artmicro,TheseHK}, diagrammatic resummations \cite{CKS2006} or {linearized} time-dependent BCS equations \cite{Annalen}.

Since Eq.~\eqref{eq:RPA} is invariant under the change of $z$ to $-z$, we {impose} {$\textrm{Re}\,z\geq0 $}. The matrix $M$ then has a branch cut for $z\in\mathcal{C}_\qq=\{E_{\kk\qq},\kk\in\mathbb{R}^3\}$, originating in {the denominator}  $z-E _ {\kk \qq} $ in (\ref{eq:Mpp}--\ref{eq:Mpm}). As such, Eq~\eqref{eq:RPA} has {at most} one solution for fixed $\qq$: {it} is real, below the continuum, 
and corresponds to the bosonic Anderson-Bogoliubov branch \cite {CKS2006}. Conversely, the collective modes we want to characterize are \textit{inside} the continuum, that is, \textit {a priori} for $\mbox {Re} \, z_\qq> \mbox{min} \, \mathcal {C}_\qq $. As in the textbook problem of one atom coupled to the electromagnetic field \cite{Cohen}, the correct way to solve Eq.~\eqref{eq:RPA} in presence of the continuum is to analytically continue the matrix $M$ through its branch cut \cite{Schmid1968}. 
This is an opportunity to recall the procedure of Nozières \cite{Nozieres1963} to analytically continue a function of the form
\be
f(z)=\int_{-\infty}^{+\infty} \dd \omega \frac{\rho(\omega)}{z-\omega},
\label{eq:fz}
\ee
analytic for $\mbox{Im}\,z\neq0$ but exhibiting a branch cut on the real axis, wherever the spectral density $ \rho $ is nonzero.
The non-analytic contribution to $M_{\sigma \sigma'} $, with $\sigma, \sigma' = \pm $, is naturally cast into this form with the spectral densities
\be
\rho_{\sigma\sigma'} (\omega,\qq)=\int \frac{\dd^3 k}{2} {W_{\kk\qq}^\sigma W_{\kk\qq}^{\sigma'}} \delta(\hbar\omega-E_{\kk\qq})
\label{eq:rho}
\ee
The analytic continuation of $f$ from upper to lower half-plane, through an interval ${[\omega_1, \omega_2 ]}$ of the branch cut where  $\rho$ is analytic, is simply
\be
f_\downarrow(z)=\begin{cases} f(z) \quad \mbox{if} \quad \mbox{Im}\, z>0 \\ 
f(z) - 2\ii\pi\rho(z) \quad \mbox{if} \quad \mbox{Im}\, z\leq0 \end{cases}
\label{eq:prescription}
\ee
where $\rho(z)$ is the analytic continuation of $\rho$ for $\mbox{Im}\, z\neq0$.
This is readily demonstrated by writing $\rho(\omega)=[\rho(\omega)-\rho(z)]+\rho(z)$ in \eqref{eq:fz} {with an energy cut-off}.

\begin{figure}
\begin{center}
\includegraphics[width=0.5\textwidth]{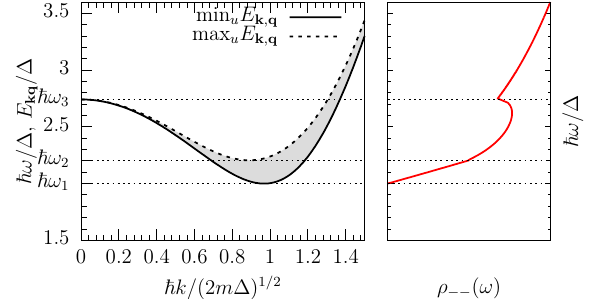}
\end{center}
\caption{\label{fig:energie} {Left: As a function of $k$, the interval between $\mbox{min}_u E _ {\kk \qq} $ (reached for $u = 0$, solid line) and $\mbox {max}_u E_ {\kk \qq}$ (reached for $u = \pm1 $, dashed line) determines an energy band (gray area) in which the resonance $\hbar\omega = E _ {\kk \qq} $ occurs for at least one value of $ u = \cos (\widehat{\kk, \qq}) $ in $ [- 1,1] $.} For fixed $\omega$, the integration interval over $k$ in \eqref{eq:rho} is read horizontally; as a function of $\omega $, its structure undergoes 3 transitions in $ \omega_1 $, $ \omega_2 $ and $ \omega_3 $, which results in angular points in the spectral density. Right: Example of $ \rho _ {--} $ (solid line). {Here, $ \mu / \Delta = 1 $ and $ \hbar q / \sqrt {2m \Delta} = 0.5 $.}}
\end{figure}

To carry out the analytic continuation of $M$, we study the function $\omega \mapsto \rho _ {\sigma \sigma '}$ on the real axis, and search for singularities. For that, we integrate over $\kk$ in \eqref{eq:rho} in a spherical frame of axis $\qq$ and use the Dirac-$\delta$ to perform the angular integration over $ u = \kk \cdot \qq / kq $. The remaining integral {over} $k$ is restricted to a domain represented on Fig.~\ref{fig:energie}, whose form depends on $\omega$.
When $\mu> 0 $ {the BCS excitation branch has its minimum in $k_0=\sqrt{2m\mu/\hbar^2}$; then, for {$q>0$ small enough} \footnote{{From some $q=q_0 < 2k_0$,  $k\mapsto\textrm{max}_u E_{\kk,\qq}$ is minimal in $k = 0$, hence $\omega_3(q)=\omega_2(q)$. For $q>2k_0$, $\omega_3(q)=\omega_2(q)=\omega_1(q)>2\Delta/\hbar$.}} the function $\omega\mapsto\rho_{\sigma\sigma'}$ has three angular points related to a configuration change of the integration domain, which divides the real axis in four distinct sectors (see Fig.~\ref{fig:energie}): (i) for $ \omega<\omega_1 = 2 \Delta/\hbar $, the resonance condition $\hbar\omega = E _ {\kk \qq} $ is never {satisfied}, so that $\rho_{\sigma\sigma'}(\omega<\omega_1)=0$,
  (ii) for $\omega_1<\omega<\omega_2$ it is reached {on an interval} $ [k_1, k_2]$,
(iii) for $\omega_2<\omega<\omega_3$, it occurs {on disjoint intervals} $ [k_1, k_1 ']$ and $[k_2', k_2] $, 
and (iv) for $\omega>\omega_3 $, { it occurs again on an interval $ [k_2', k_2] $}.

\textit{Numerical study at arbitrary $q$ --} 
We find a solution $z_\qq=\hbar\omega_\qq-\ii\hbar{\Gamma_\qq}/{2}$ to Eq.~\eqref{eq:RPA} in the analytic continuation through the {sector} $[ \omega_1, \omega_2 ]$ {(see the schematic on Fig.~\ref{fig:schema})}, which we identify as the energy of the sought collective mode. {In this sector, we express the spectral functions in terms of first and second kind complete elliptic integrals} \footnote{{If $\ch\Omega=\hbar\omega/2\Delta$ and $\hbar=2m=1$, $\rho_{++}(\omega)=\frac{\pi\Delta}{q}E(\ii\sh\Omega)$, $\rho_{--}(\omega)=\rho_{++}(\omega)-\frac{\pi\Delta}{q}K(\ii\sh\Omega)$, $\rho_{+-}(\omega)=0$.}}. 
\begin{figure}[htb]
\begin{center}
\includegraphics[width=0.5\textwidth]{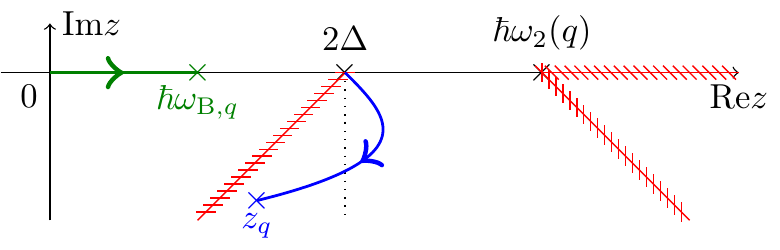}
\vspace{-1cm}
\end{center}
\caption{Trajectories of the {pair-breaking} collective branch 
(blue curve) and Bogoliubov-Anderson branch
(green line) as functions of $q$ in the complex plane. The first one is revealed only after analytic continuation, hence the deformed branch cut (striped red lines) in the lower half-plane. \label{fig:schema}}
\end{figure}

The dispersion relation $q \mapsto \omega_ \qq$ is represented on Fig.~\ref{fig:disp} for pairing strengths $ \mu/\Delta = 1/10 $, $ 5 $ and $ 100 $ ($ 1 / k _ {\rm F} a \simeq0.5 $, $ {-1.1} $ and $ -3.0 $ in Fermi gases {with Fermi wavenumber $k_{\rm F}$ and scattering length $a$}). Departing quadratically from its limit $2\Delta$ in $q = 0$, the branch goes through a maximum of height proportional to $\Delta$ {and location of order the inverse of the pair radius $\xi\approx \hbar^2k_0/m\Delta$} at weak coupling $ \Delta \ll \mu $, then {dips below $2\Delta$}. In the strong coupling regime $\Delta>\mu$, the domain where the energy of the branch is greater than $ 2 \Delta $ shrinks, until its disappearance for $ \mu / \Delta \simeq 0.8267 $. Conversely, the damping rate $ \Gamma_\qq $ is a strictly increasing function of $q$, also {starting quadratically} from its zero limit in $ q=0$. This is in direct contrast with the commonly accepted prediction in the literature of a damping rate linear in $q$ \cite{Varma1982}. {The fact that our solution travels far away from the initial branch cut underlines the non-perturbative nature of our analytic continuation: there is no unperturbed solution on the real axis from which $\textrm{Im}z_\qq$ could be deduced {from Fermi's golden rule}.}

The branch disappears in $ q = 2k_0 $ (hence before the Bogoliubov-Anderson branch hits the continuum \cite{CKS2006}) when the interval $ [\omega_1, \omega_2] $ through which our analytic continuation passes reduces to a point.
Last, we exclude {the existence of} a branch of energy above $ 2\sqrt {\Delta ^ 2 + \mu ^ 2}$ {(twice the gap)} in the BEC regime where $ \mu <0 $ and where the three singularities $\omega_i$ of $ \rho_ {\sigma \sigma '} $ gather.
\begin{figure}
\begin{center}
\includegraphics[width=0.5\textwidth]{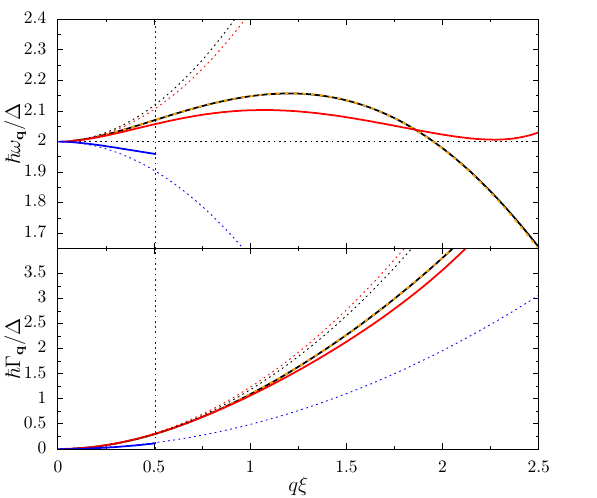}
\end{center}
\caption{\label{fig:disp} 
Frequency (top) and damping rate (bottom) of the {pair-breaking} collective mode as functions of $q$ for $ \mu / \Delta = 100 $ (black solid curve), $ \mu / \Delta = 5 $ (red solid curve) and $ \mu / \Delta = 0.1 $ (blue solid curve, {disappears in $2k_0\xi\simeq0.51$}) {as functions of $q$ in units of the inverse pair size $\xi$ \cite{Strinati1998}. Dashed curves: the same for $ \mu / \Delta = 100$ omitting the amplitude-phase coupling $M_{+-}$.} Dotted curves: {low-$q$ quadratic} behavior obtained analytically from Eqs.~\eqref{eq:reparam}-\eqref{eq:apres}.}
\end{figure}

\textit{Long wavelength limit --} In this limit, we obtain several analytical results that corroborate our numerical study. We deal separately with the singular case $q = 0$, where the matrix $ M(z,\qq = 0)$ is expressible
in terms of first and third kind complete elliptic {integrals} $K(k) $ and $ \Pi (n, k) $~\cite{Gradshteyn}~\footnote{The $\omega$-integral giving the dimensionless $\tilde{M}_{\sigma\sigma'}= {M}_{\sigma\sigma'}\Delta(\hbar^2/2m\Delta)^{3/2}$ for $q=0$ in \eqref{eq:fz} is reduced to elliptic integrals \cite{Gradshteyn} by the change of variable $\hbar \omega = \Delta (x^2+1/x^2), x \in [0,1]$.}:
\bea
\mbox{th} s \tilde{M}_{++}(z,0) \!\!&=&\!\! \frac{\tilde{M}_{--}(z,0)}{\mbox{th} s} = - \pi(2 \eee^l)^{1/2} [F(s)-F(-s)] \notag \\
\tilde M_{+-}(z,0) \!\!&=&\!\! - \pi(2 \eee^l)^{1/2} [F(s) + F(-s)] \label{eq:Mpmq0}
\eea
with $ l=\textrm{argsh}(\mu/\Delta) $, $ s = \textrm{argch}(z/2\Delta)$, {and}
\be
F(s)\!\!=\!\!(\sh l +\sh s)[\Pi(\eee^{l+s},\ii\eee^l)-\Pi(-\eee^{l-s},\ii\eee^l)]+K(\ii\eee^l)\ch s
\ee
{Eq.~\eqref{eq:RPA} then reads simply $F(s) F(-s)=0$.} Even after analytic continuation \footnote{The branch cut $[2\Delta,+\infty[$ 
in $z$ for $\mu> 0 $ translates into a branch {cut} $[0,+\infty[$ in $s$.
Thus $F(s)-F(-s)$ has the nonzero limit $\ii\pi(1+\sqrt{1+\Delta^2/\mu^2})^{-1/2}/\sqrt{2}$ when $s\to0$ with $\im s>0$.
} this equation has no solution besides {$s=\ii\pi/2$} ($z = 0$, {the} starting point of the Anderson-Bogoliubov branch); in particular $ F (s) $ has a finite nonzero limit when $z\to2\Delta$ ($s\to0$) with $\mbox{Im}s> 0 $.
Thus, the threshold of the pair-breaking continuum $\omega = 2 \Delta/\hbar $ is not a solution of the RPA equation \eqref{eq:RPA} in $q=0$  \cite{Matera2010}, and not a pole of the response functions.
This is why, as understood by Refs.~\cite{Kogan1973,Orbach1981,Altshuler2006,Gurarie2009,Stringari2012,Foster2015}, the ``Higgs'' oscillations 
at this frequency are not sinusoidal as $\cos(2 \Delta t/\hbar + \phi)$ but subject to a power-law damping as $\cos(2 \Delta t/\hbar + \phi)/t^\alpha$, {$\alpha>0$}.

For small but nonzero $q$, and $\mu> 0 $, the resonance sector between $ \hbar \omega_1 = 2 \Delta $ and $\hbar\omega_2=2\Delta+\mu {\hbar^2q^2}/{2m\Delta}+O(q^4)$ in Fig.~\ref{fig:energie} has a width $O(q^2)$ in energy, and $O(q)$ in the {wavenumber} $ k $ around the minimum {location} $k_0$ of the BCS branch. We then set
\be
z_\qq=2\Delta+\zeta\frac{\hbar^2q^2}{4m^*} + O(q^3) \quad \mbox{and} \quad k=k_0+Kq
\label{eq:reparam}
\ee
with $m^*=m\Delta/2\mu$ the effective mass of the BCS branch minimum. We thus focus on the wavevector domain where the denominator in (\ref{eq:Mpp},\ref{eq:Mpm}) is of order $q^2$:
\be
z-E_{\kk\qq} = z-2\Delta -\frac{\hbar^2 q^2}{{m^*}} (K^2 +u^2/4) + O(q^3)
\ee
Now, using the expansions of the numerator amplitudes $W_{\kk\qq}^+\sim1$ and $W_{\kk\qq}^-\sim\hbar^2 k_0 q K/m\Delta$,
{and performing the integral over the angular variable $u$ before that over $K$}
we obtain the analytic expressions for $\mbox{Im}\, z>0$:
\bea
\!\!\!\!\!\!\!\!\!\tilde{M}_{++}(z,\qq) \!\!\!&\underset{q\to0}{\sim}&\!\!\! - \frac{\ii\pi^2 (2m\Delta)^{1/2}}{\hbar q}\asin \frac{1}{\sqrt{\zeta}}  \label{eq:devMpp}\\
\!\!\!\!\!\!\!\!\!\tilde{M}_{--}(z,\qq) \!\!\!&\underset{q\to0}{\sim}&\!\!\! -\frac{\ii\pi^2 \mu \hbar q}{(8m\Delta^3)^{1/2}}\bbcro{\sqrt{\zeta-1}+\zeta\asin \frac{1}{\sqrt{\zeta}}}  \label{eq:devMmm} 
\eea
Since the divergence of $ M _ {++} $ of order $1 / q$ is compensated by the {suppression} of $M_{--} $ linear in $ q $, the finite nonzero limit {\eqref{eq:Mpmq0} of $ M_{+ -} $ in $q = 0,\,\, \hbar\omega=2\Delta $ suffices}.
Inserting expressions (\ref{eq:Mpmq0},\ref{eq:devMpp},\ref{eq:devMmm}) in the RPA equation \eqref{eq:RPA} and analytically continuing the product $M_{++} M_{--} $ through its branch cut $ [0,1] $ in $\zeta$ (corresponding to the segment $[\hbar \omega_1, \hbar \omega_2] $ in $ z $) with the substitutions $\asin 1/\sqrt{\zeta} \to \pi-\asin 1/\sqrt{\zeta}$ and $\sqrt{\zeta-1}\to -\sqrt{\zeta-1}$, we obtain an explicit yet transcendental equation on $\zeta$:
\begin{multline}
\bbcro{\pi-\asin \frac{1}{\sqrt{\zeta}}} \bbcro{\bb{\pi-\asin \frac{1}{\sqrt{\zeta}}}\zeta- \sqrt{\zeta-1}} \\ + \frac{2}{\pi^4\mu} \bb{\frac{\hbar^2}{2m}}^{3} {M}_{+-}^2(2\Delta,0)=0
\label{eq:apres}
\end{multline}
The continuation is for the entire lower half-plane, including $\mbox{Re}\, z<2\Delta$ ($\mbox{Re}\,\zeta<0$). The unique solution of Eq.~\eqref{eq:apres} shown in Fig.~\ref{fig:z} faithfully reproduces the coefficient of $q^2$ in Fig.~\ref{fig:disp}. The real part changes sign for {$\mu/\Delta\simeq0.8267$}, which confirms that the branch is below $ 2 \Delta $ at strong coupling.

To understand the disappearance of the branch at $q=0$, we calculate the matrix-residue of 
$M\!\downarrow\!(z,\qq)^{-1}$ at $z_\qq$ and find that it vanishes linearly: it becomes proportional to the amplitude-channel projector $\begin{pmatrix}0&0\\0&1\end{pmatrix}$ with a factor
\be
Z_\qq\underset{q\to 0}{\sim} \frac{\ii \hbar^4 q}{2 m^2 \pi^2} \frac{\pi-\asin\frac{1}{\sqrt{\zeta}}}{(\pi-\asin\frac{1}{\sqrt{\zeta}})^2+\frac{(\pi-\asin\frac{1}{\sqrt{\zeta}})\zeta-\sqrt{\zeta-1}}{2\zeta\sqrt{\zeta-1}}}
\ee
This results from applying $\frac{\dd}{\dd z}\propto q^{-2}\frac{\dd}{\dd\zeta}$ to Eqs.~(\ref{eq:devMpp},\ref{eq:devMmm}). {$Z_\qq$ is the weight of
the collective mode above the continuum background; its suppression in $q=0$ means that the many-body response function can no longer be interpreted in terms of a quasiparticle on an incoherent background.}


At weak coupling ($\mu/\Delta\to +\infty$), $ M_{+-} $ tends to zero because of the {antisymmetry
$k \leftrightarrow 2k_0-k $ about the Fermi surface,} {valid for $(k-k_0)\xi=O(1)$}. The RPA equation reduces to {$M_{++} M _ {--} = 0$} for {$ q\xi ={O}(1) $}, {and Eq.~\eqref{eq:apres} to its $\zeta$-dependent first line.} The {pair-breaking} collective mode is then a pure amplitude mode (a root of $M_{\downarrow--}$), while the phononic phase mode solves $M_{++}=0$ \footnote{{Although $M_{--}(2 \Delta,0)=0$, there is no amplitude mode in $q=0$ because $1/M_{--}(z,0)$ is not meromorphic.}}. {Its quadratic dispersion relation}
\be
z_\qq \stackrel[\mu/\Delta\to+\infty]{q \to 0}{\simeq} 2\Delta { + (0.2369-0.2956\ii) \frac{\hbar^2q^2}{4m^*} }
\label{couplfaible}
\ee
{contradicts Ref.~\cite{Varma1982} (even $\textrm{Re}\,\zeta$ differs from the value {$1/3$} of \cite{Varma1982}), but confirms \cite{Popov1976}.}

Our calculation shows the limits of the analogy with Higgs modes in field theory: although it is also a gapped amplitude mode at weak coupling, the collective mode, here immersed in a continuum, is obtained only after a non-perturbative treatment of the coupling to fermionic degrees of freedom; impossible therefore to obtain it reliably from a low-energy ($\hbar\omega\ll2\Delta$) effective action as suggested sometimes \cite{Varma2015,Zhang2016}.

\begin{figure}
\begin{center}
\includegraphics[width=0.48\textwidth]{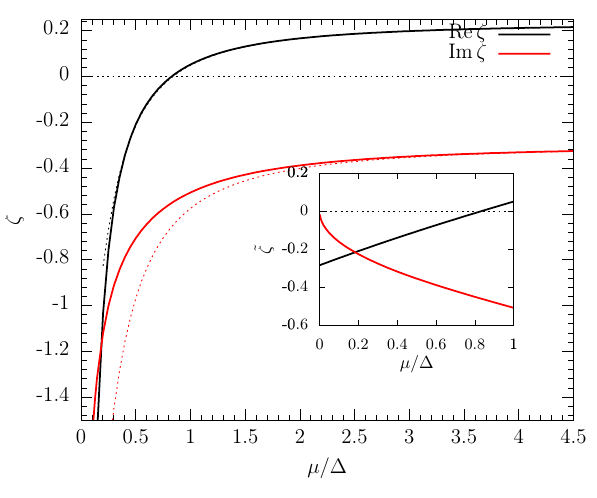}
\end{center}
\vspace{-0.5cm}

\caption{\label{fig:z} 
Real and imaginary parts ({black and red} solid curves) of the dimensionless coefficient $\zeta$
 of $q^2$ in the energy $z_\qq$ of the {pair-breaking collective} mode as functions of $ \mu/\Delta $. Dashed curves: weak coupling expansion {$\zeta=\zeta_0-\frac{2 \zeta_0^2}{\zeta_0-1} \left(\frac{\Delta}{\pi\mu}\right)^2 \ln^2 \frac{\Delta}{8\mu \eee} + \ldots$ with $\zeta_0\simeq0.2369-0.2956\ii$}. Inset: rescaled coefficient $ \tilde{\zeta} = \zeta \mu / \Delta = \zeta m / 2m ^ * $ {admitting} the finite real limit $\tilde{\zeta}_{\infty}=-16 K^2(\ii)/\pi^4\simeq-0.2823$ at strong coupling $\mu/\Delta\to0^+ $, its imaginary part tending to zero like $-12 K(\ii) (\mu/\Delta)^{1/2}/\pi^3$.}
\end{figure}

\textit {Observability in response functions --} At low $q$, the {pair-breaking} collective mode is {weakly damped, a favorable condition}. At weak coupling, as shown in Fig.~\ref{fig:reponse}, there indeed appears in the response function of the order-parameter amplitude a smooth peak, whose position, width and height are remarkably predicted by the branch obtained in the analytic continuation. At strong enough coupling (blue curve in Fig.~5b), the smooth resonance peak disappears and there remains a sharp one (with a vertical tangent), whose maximum is at $\omega=2\Delta/\hbar$ even for $q\neq0$. Qualitatively, this indicates that the collective frequency $\omega_\qq$ is below $2\Delta/\hbar$ such that there is no complex resonance in the interval $[2\Delta/\hbar, \omega_2]$ where our analytic continuation is meaningful.

The amplitude response function ($ |M_{++}/\textrm{det}M(\omega+\ii0^+,\qq)|^2$, or $1/|M_{--}(\omega+\ii0^+,\qq)|^2$ at weak coupling), unlike the more commonly measured density-density response \cite{Vale2017}, is sensitive to the pair-breaking collective mode even at weak coupling. In cold gases, the order-parameter amplitude can be excited by Feshbach-modulation of the interaction strength, and measured by spatially resolved interferometry \cite{CarusottoCastin2005}.
Physically, Fig.~\ref{fig:reponse} shows that the system absorbs energy from modulations of the pairing strength $|\Delta|$ at frequencies $\omega>2\Delta/\hbar$ more efficiently when $\omega$ is close to $\omega_\qq$. This resonance is broadened because the absorbed energy is dissipated by breaking pairs into unpaired fermions of wavevectors $\qq/2\pm\kk$.

\begin{figure}
\begin{center}
\includegraphics[width=0.48\textwidth]{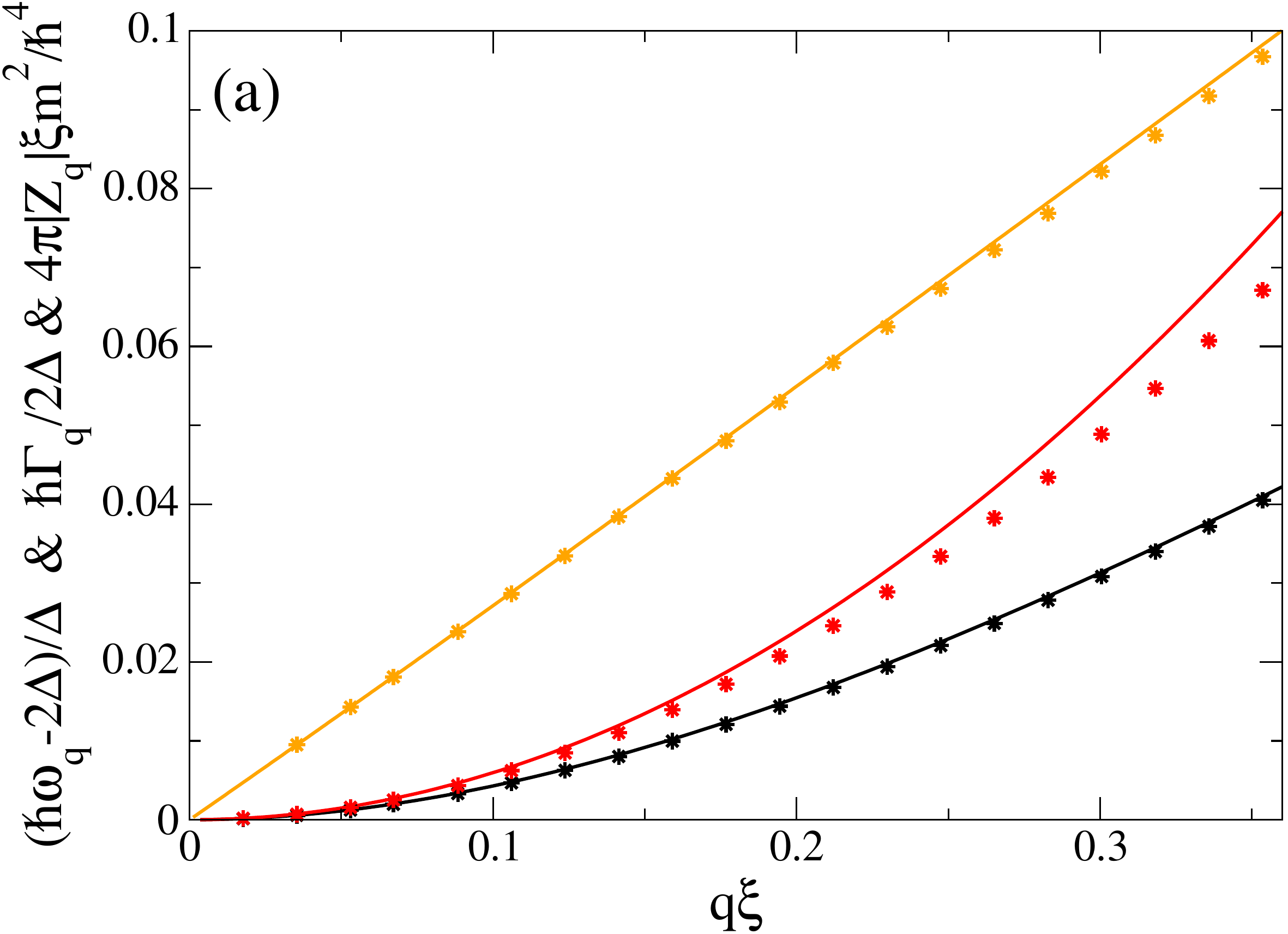}

\includegraphics[width=0.48\textwidth]{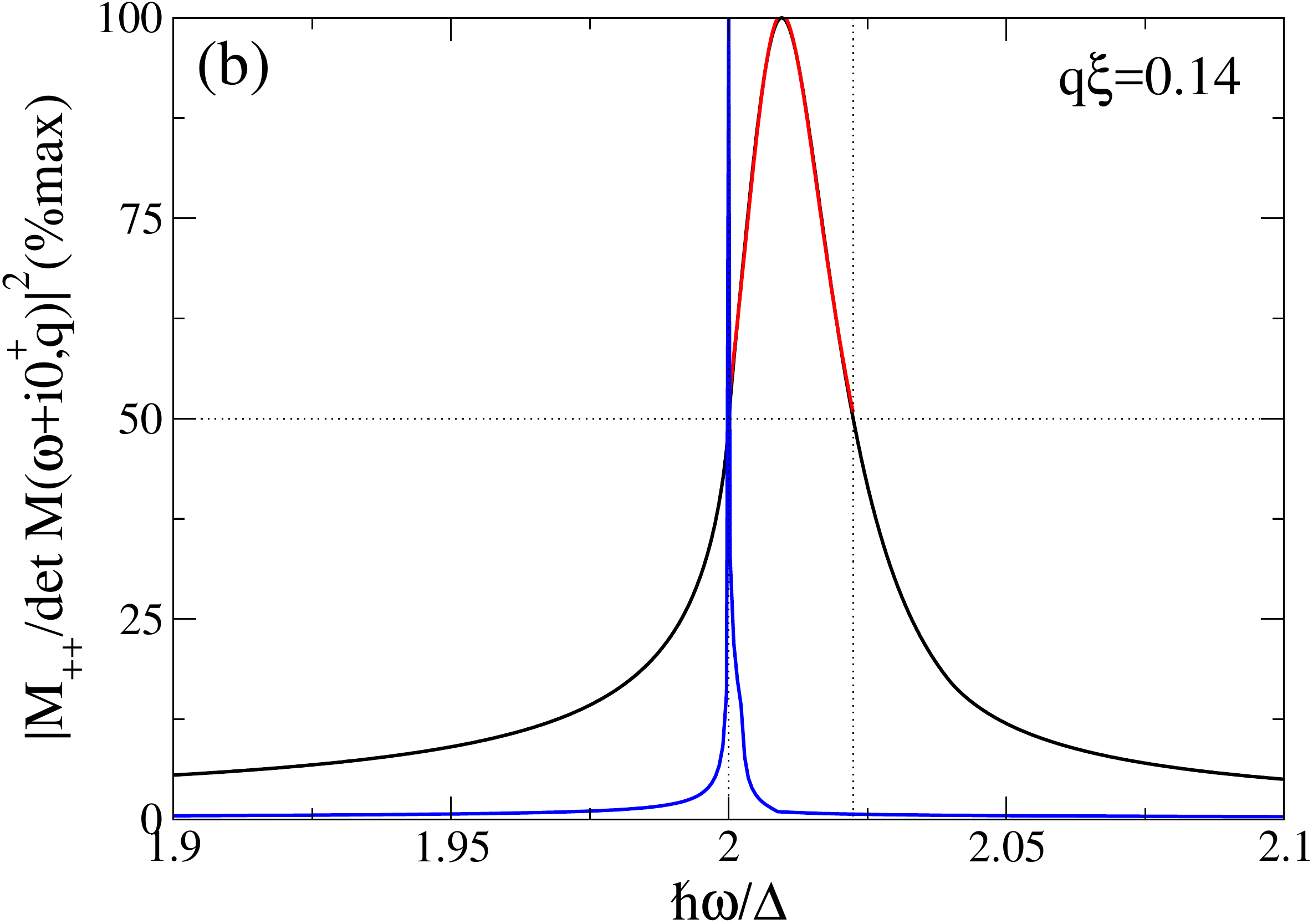}
\end{center}
\vspace{-0.5cm}
\caption{\label{fig:reponse} 
{$(a)$ At weak coupling ($\Delta/\mu\to0$), frequency displacement $\omega_\qq-2\Delta/\hbar$, damping rate $\Gamma_\qq$ and residue $Z_\qq$ of the pair-breaking collective mode (black, red, orange solid lines) compared to the values (stars) extracted by fitting the amplitude response function $1/|M_{--}(\omega+\ii0^+,\qq)|^2$ {(black curve of $(b)$) by} the function $|C+{Z_\qq}^{\rm fit}/(\omega-\omega_\qq^{\rm fit}+\ii\Gamma_\qq^{\rm fit}/2)|^2$ {(red curve of $(b)$)} describing a resonance on a flat background $C$. {Blue curve of $(b)$: amplitude response function $|M_{++}/\textrm{det}M(\omega+\ii0^+,\qq)|^2$ at strong coupling ($\Delta/\mu=10$) exhibiting only a sharp peak at $2\Delta/\hbar$.}}
}
\end{figure}

\textit{Conclusion --} We have established on solid theoretical foundations the existence of a collective branch inside the pair-breaking continuum of BCS superconductors and superfluid Fermi gases, and we have fully characterized its dispersion relation and damping rate, {including in the strong coupling regime where it is a mixture of amplitude and phase fluctuations.} We thus give a complete answer to an old condensed-matter problem. The branch appears clearly {in the order-parameter response function which can be measured in cold atomic gases}.

\begin{acknowledgements}

\end{acknowledgements}

\bibliography{/Users/rox/Documents/biblio}

\providecommand*\hyphen{-}
\begin{thebibliography}{10}
\expandafter\ifx\csname fonteauteurs\endcsname\relax
\def\fonteauteurs{\scshape}\fi
\makeatother

\bibitem{Anderson1958}
P.W. \bgroup\fonteauteurs\bgroup Anderson\egroup\egroup{} :
\newblock {Random-Phase Approximation in the Theory of Superconductivity}.
\newblock {\em Phys. Rev.}, 112\string:\penalty500\relax 1900--1916, 1958.

\bibitem{Strinati1998}
M.~\bgroup\fonteauteurs\bgroup Marini\egroup\egroup{},
  F.~\bgroup\fonteauteurs\bgroup Pistolesi\egroup\egroup{} et G.C.
  \bgroup\fonteauteurs\bgroup Strinati\egroup\egroup{} :
\newblock {Evolution from BCS superconductivity to Bose condensation: analytic
  results for the crossover in three dimensions}.
\newblock {\em European Physical Journal B}, 1\string:\penalty500\relax
  151--159, 1998.

\bibitem{CKS2006}
R.~\bgroup\fonteauteurs\bgroup Combescot\egroup\egroup{}, M.~Yu.
  \bgroup\fonteauteurs\bgroup Kagan\egroup\egroup{} et
  S.~\bgroup\fonteauteurs\bgroup Stringari\egroup\egroup{} :
\newblock {Collective mode of homogeneous superfluid Fermi gases in the BEC-BCS
  crossover}.
\newblock {\em Phys. Rev. A}, 74\string:\penalty500\relax 042717, octobre 2006.

\bibitem{Tachiki1997}
K.~\bgroup\fonteauteurs\bgroup Kadowaki\egroup\egroup{},
  I.~\bgroup\fonteauteurs\bgroup Kakeya\egroup\egroup{}, M.~B.
  \bgroup\fonteauteurs\bgroup Gaifullin\egroup\egroup{},
  T.~\bgroup\fonteauteurs\bgroup Mochiku\egroup\egroup{},
  S.~\bgroup\fonteauteurs\bgroup Takahashi\egroup\egroup{},
  T.~\bgroup\fonteauteurs\bgroup Koyama\egroup\egroup{} et
  M.~\bgroup\fonteauteurs\bgroup Tachiki\egroup\egroup{} :
\newblock {Longitudinal Josephson-plasma excitation in
  ${\mathrm{Bi}}_{2}{\mathrm{Sr}}_{2}{\mathrm{CaCu}}_{2}{\mathrm{O}}_{8+\ensuremath{\delta}}:$
  Direct observation of the Nambu-Goldstone mode in a superconductor}.
\newblock {\em Phys. Rev. B}, 56\string:\penalty500\relax 5617--5621, septembre
  1997.

\bibitem{Thomas2007}
J.~\bgroup\fonteauteurs\bgroup Joseph\egroup\egroup{},
  B.~\bgroup\fonteauteurs\bgroup Clancy\egroup\egroup{},
  L.~\bgroup\fonteauteurs\bgroup Luo\egroup\egroup{},
  J.~\bgroup\fonteauteurs\bgroup Kinast\egroup\egroup{},
  A.~\bgroup\fonteauteurs\bgroup Turlapov\egroup\egroup{} et J.~E.
  \bgroup\fonteauteurs\bgroup Thomas\egroup\egroup{} :
\newblock {Measurement of Sound Velocity in a Fermi Gas near a Feshbach
  Resonance}.
\newblock {\em Phys. Rev. Lett.}, 98\string:\penalty500\relax 170401, avril
  2007.

\bibitem{Vale2017}
Sascha \bgroup\fonteauteurs\bgroup Hoinka\egroup\egroup{}, Paul
  \bgroup\fonteauteurs\bgroup Dyke\egroup\egroup{}, Marcus~G.
  \bgroup\fonteauteurs\bgroup Lingham\egroup\egroup{}, Jami~J.
  \bgroup\fonteauteurs\bgroup Kinnunen\egroup\egroup{}, Georg~M.
  \bgroup\fonteauteurs\bgroup Bruun\egroup\egroup{} et Chris~J.
  \bgroup\fonteauteurs\bgroup Vale\egroup\egroup{} :
\newblock {Goldstone mode and pair-breaking excitations in atomic Fermi
  superfluids}.
\newblock {\em Nature Physics}, 13\string:\penalty500\relax 943--946, juin
  2017.

\bibitem{Varma2015}
David \bgroup\fonteauteurs\bgroup Pekker\egroup\egroup{} et C.M.
  \bgroup\fonteauteurs\bgroup Varma\egroup\egroup{} :
\newblock {Amplitude/Higgs Modes in Condensed Matter Physics}.
\newblock {\em Annual Review of Condensed Matter Physics},
  6(1)\string:\penalty500\relax 269--297, 2015.

\bibitem{Schmid1968}
Albert \bgroup\fonteauteurs\bgroup Schmid\egroup\egroup{} :
\newblock {The approach to equilibrium in a pure superconductor. The relaxation
  of the Cooper pair density}.
\newblock {\em Physik der kondensierten Materie}, 8(2)\string:\penalty500\relax
  129--140, novembre 1968.

\bibitem{Orbach1981}
I.~O. \bgroup\fonteauteurs\bgroup Kulik\egroup\egroup{}, Ora
  \bgroup\fonteauteurs\bgroup Entin-Wohlman\egroup\egroup{} et
  R.~\bgroup\fonteauteurs\bgroup Orbach\egroup\egroup{} :
\newblock {Pair susceptibility and mode propagation in superconductors: A
  microscopic approach}.
\newblock {\em Journal of Low Temperature Physics},
  43(5)\string:\penalty500\relax 591--620, juin 1981.

\bibitem{Popov1976}
V.~A. \bgroup\fonteauteurs\bgroup Andrianov\egroup\egroup{} et V.~N.
  \bgroup\fonteauteurs\bgroup Popov\egroup\egroup{} :
\newblock {Gidrodinamičeskoe dejstvie i Boze-spektr sverhtekučih
  Fermi-sistem}.
\newblock {\em Teoreticheskaya i Matematicheskaya Fizika},
  28\string:\penalty500\relax 341--352, 1976.
\newblock [English translation: Theoretical and Mathematical Physics, 1976,
  28:3, 829–837].

\bibitem{Popov1987-III13}
V.~N. \bgroup\fonteauteurs\bgroup Popov\egroup\egroup{} :
\newblock {B}ose spectrum of superfluid {F}ermi gases.
\newblock \emph{In} {\em {F}unctional {I}ntegral and {C}ollective
  {E}xcitations}, chapitre III, section 13. Cambridge University Press,
  Cambridge, 1987.

\bibitem{Varma1982}
P.~B. \bgroup\fonteauteurs\bgroup Littlewood\egroup\egroup{} et C.~M.
  \bgroup\fonteauteurs\bgroup Varma\egroup\egroup{} :
\newblock {Amplitude collective modes in superconductors and their coupling to
  charge-density waves}.
\newblock {\em Phys. Rev. B}, 26\string:\penalty500\relax 4883--4893, novembre
  1982.

\bibitem{Benfatto2016}
T.~\bgroup\fonteauteurs\bgroup Cea\egroup\egroup{},
  C.~\bgroup\fonteauteurs\bgroup Castellani\egroup\egroup{} et
  L.~\bgroup\fonteauteurs\bgroup Benfatto\egroup\egroup{} :
\newblock {Nonlinear optical effects and third-harmonic generation in
  superconductors: Cooper pairs versus Higgs mode contribution}.
\newblock {\em Phys. Rev. B}, 93\string:\penalty500\relax 180507, mai 2016.

\bibitem{Stringari2012}
R.~G. \bgroup\fonteauteurs\bgroup Scott\egroup\egroup{},
  F.~\bgroup\fonteauteurs\bgroup Dalfovo\egroup\egroup{}, L.~P.
  \bgroup\fonteauteurs\bgroup Pitaevskii\egroup\egroup{} et
  S.~\bgroup\fonteauteurs\bgroup Stringari\egroup\egroup{} :
\newblock {Rapid ramps across the BEC-BCS crossover: A route to measuring the
  superfluid gap}.
\newblock {\em Phys. Rev. A}, 86\string:\penalty500\relax 053604, novembre
  2012.

\bibitem{Kogan1973}
A.F. \bgroup\fonteauteurs\bgroup Volkov\egroup\egroup{} et Ch.~M.
  \bgroup\fonteauteurs\bgroup Kogan\egroup\egroup{} :
\newblock Collisionless relaxation of the energy gap in superconductors.
\newblock {\em Zh. Eksp. Teor. Fiz.}, 65\string:\penalty500\relax 2038, 1973.

\bibitem{Altshuler2006}
Emil~A. \bgroup\fonteauteurs\bgroup Yuzbashyan\egroup\egroup{}, Oleksandr
  \bgroup\fonteauteurs\bgroup Tsyplyatyev\egroup\egroup{} et Boris~L.
  \bgroup\fonteauteurs\bgroup Altshuler\egroup\egroup{} :
\newblock {Relaxation and Persistent Oscillations of the Order Parameter in
  Fermionic Condensates}.
\newblock {\em Phys. Rev. Lett.}, 96\string:\penalty500\relax 097005, mars
  2006.

\bibitem{Gurarie2009}
V.~\bgroup\fonteauteurs\bgroup Gurarie\egroup\egroup{} :
\newblock {Nonequilibrium Dynamics of Weakly and Strongly Paired
  Superconductors}.
\newblock {\em Phys. Rev. Lett.}, 103\string:\penalty500\relax 075301, ao\^ut
  2009.

\bibitem{Foster2015}
E.~A. \bgroup\fonteauteurs\bgroup Yuzbashyan\egroup\egroup{},
  M.~\bgroup\fonteauteurs\bgroup Dzero\egroup\egroup{},
  V.~\bgroup\fonteauteurs\bgroup Gurarie\egroup\egroup{} et M.~S.
  \bgroup\fonteauteurs\bgroup Foster\egroup\egroup{} :
\newblock {Quantum quench phase diagrams of an $s$-wave BCS-BEC condensate}.
\newblock {\em Phys. Rev. A}, 91\string:\penalty500\relax 033628, mars 2015.

\bibitem{Klein1980}
R.~\bgroup\fonteauteurs\bgroup Sooryakumar\egroup\egroup{} et M.~V.
  \bgroup\fonteauteurs\bgroup Klein\egroup\egroup{} :
\newblock {Raman Scattering by Superconducting-Gap Excitations and Their
  Coupling to Charge-Density Waves}.
\newblock {\em Phys. Rev. Lett.}, 45\string:\penalty500\relax 660--662, ao\^ut
  1980.

\bibitem{Shimano2013}
Ryusuke \bgroup\fonteauteurs\bgroup Matsunaga\egroup\egroup{}, Yuki~I.
  \bgroup\fonteauteurs\bgroup Hamada\egroup\egroup{}, Kazumasa
  \bgroup\fonteauteurs\bgroup Makise\egroup\egroup{}, Yoshinori
  \bgroup\fonteauteurs\bgroup Uzawa\egroup\egroup{}, Hirotaka
  \bgroup\fonteauteurs\bgroup Terai\egroup\egroup{}, Zhen
  \bgroup\fonteauteurs\bgroup Wang\egroup\egroup{} et Ryo
  \bgroup\fonteauteurs\bgroup Shimano\egroup\egroup{} :
\newblock {Higgs Amplitude Mode in the BCS Superconductors
  ${\mathrm{Nb}}_{1\mathrm{\text{\ensuremath{-}}}x}{\mathrm{Ti}}_{x}\mathbf{N}$
  Induced by Terahertz Pulse Excitation}.
\newblock {\em Phys. Rev. Lett.}, 111\string:\penalty500\relax 057002, juillet
  2013.

\bibitem{Sacuto2014}
M.-A. \bgroup\fonteauteurs\bgroup M\'easson\egroup\egroup{},
  Y.~\bgroup\fonteauteurs\bgroup Gallais\egroup\egroup{},
  M.~\bgroup\fonteauteurs\bgroup Cazayous\egroup\egroup{},
  B.~\bgroup\fonteauteurs\bgroup Clair\egroup\egroup{},
  P.~\bgroup\fonteauteurs\bgroup Rodi\`ere\egroup\egroup{},
  L.~\bgroup\fonteauteurs\bgroup Cario\egroup\egroup{} et
  A.~\bgroup\fonteauteurs\bgroup Sacuto\egroup\egroup{} :
\newblock {Amplitude Higgs mode in the $2H\ensuremath{-}{\text{NbSe}}_{2}$
  superconductor}.
\newblock {\em Phys. Rev. B}, 89\string:\penalty500\relax 060503, f\'evrier
  2014.

\bibitem{Devereaux2015}
A.~F. \bgroup\fonteauteurs\bgroup Kemper\egroup\egroup{}, M.~A.
  \bgroup\fonteauteurs\bgroup Sentef\egroup\egroup{},
  B.~\bgroup\fonteauteurs\bgroup Moritz\egroup\egroup{}, J.~K.
  \bgroup\fonteauteurs\bgroup Freericks\egroup\egroup{} et T.~P.
  \bgroup\fonteauteurs\bgroup Devereaux\egroup\egroup{} :
\newblock {Direct observation of Higgs mode oscillations in the pump-probe
  photoemission spectra of electron-phonon mediated superconductors}.
\newblock {\em Phys. Rev. B}, 92\string:\penalty500\relax 224517, d\'ecembre
  2015.

\bibitem{Koehl2018}
A.~\bgroup\fonteauteurs\bgroup Behrle\egroup\egroup{},
  T.~\bgroup\fonteauteurs\bgroup Harrison\egroup\egroup{},
  J.~\bgroup\fonteauteurs\bgroup Kombe\egroup\egroup{},
  K.~\bgroup\fonteauteurs\bgroup Gao\egroup\egroup{},
  M.~\bgroup\fonteauteurs\bgroup Link\egroup\egroup{}, J.~S.
  \bgroup\fonteauteurs\bgroup Bernier\egroup\egroup{},
  C.~\bgroup\fonteauteurs\bgroup Kollath\egroup\egroup{} et
  M.~\bgroup\fonteauteurs\bgroup K{\"o}hl\egroup\egroup{} :
\newblock {Higgs mode in a strongly interacting fermionic superfluid}.
\newblock {\em Nature Physics}, 2018.

\bibitem{Randeria1997}
Jan~R. \bgroup\fonteauteurs\bgroup Engelbrecht\egroup\egroup{}, Mohit
  \bgroup\fonteauteurs\bgroup Randeria\egroup\egroup{} et C.~A.~R. S\'a~de
  \bgroup\fonteauteurs\bgroup Melo\egroup\egroup{} :
\newblock {BCS to Bose crossover: Broken-symmetry state}.
\newblock {\em Phys. Rev. B}, 55\string:\penalty500\relax 15153--15156, juin
  1997.

\bibitem{Randeria2008}
Roberto~B. \bgroup\fonteauteurs\bgroup Diener\egroup\egroup{}, Rajdeep
  \bgroup\fonteauteurs\bgroup Sensarma\egroup\egroup{} et Mohit
  \bgroup\fonteauteurs\bgroup Randeria\egroup\egroup{} :
\newblock {Quantum fluctuations in the superfluid state of the BCS-BEC
  crossover}.
\newblock {\em Phys. Rev. A}, 77\string:\penalty500\relax 023626, f\'evrier
  2008.

\bibitem{artmicro}
Hadrien \bgroup\fonteauteurs\bgroup Kurkjian\egroup\egroup{} et Jacques
  \bgroup\fonteauteurs\bgroup Tempere\egroup\egroup{} :
\newblock {Absorption and emission of a collective excitation by a fermionic
  quasiparticle in a Fermi superfluid}.
\newblock {\em New Journal of Physics}, 19(11)\string:\penalty500\relax 113045,
  2017.

\bibitem{TheseHK}
H.~\bgroup\fonteauteurs\bgroup Kurkjian\egroup\egroup{} :
\newblock {\em {Cohérence}, brouillage et dynamique de phase dans un condensat
  de paires de fermions}.
\newblock Th\`ese de doctorat, \'Ecole Normale Supérieure, Paris, 2016.

\bibitem{Annalen}
H.~\bgroup\fonteauteurs\bgroup Kurkjian\egroup\egroup{},
  Y.~\bgroup\fonteauteurs\bgroup Castin\egroup\egroup{} et
  A.~\bgroup\fonteauteurs\bgroup Sinatra\egroup\egroup{} :
\newblock {Three-Phonon and Four-Phonon Interaction Processes in a
  Pair-Condensed Fermi Gas}.
\newblock {\em Annalen der Physik}, 529(9)\string:\penalty500\relax 1600352,
  2017.

\bibitem{Cohen}
C.~\bgroup\fonteauteurs\bgroup Cohen-Tannoudji\egroup\egroup{},
  J.~\bgroup\fonteauteurs\bgroup Dupont-Roc\egroup\egroup{} et
  G.~\bgroup\fonteauteurs\bgroup Grynberg\egroup\egroup{} :
\newblock {\em {Processus d'interaction entre photons et atomes}}.
\newblock InterEditions et \'Editions du {CNRS}, Paris, 1988.

\bibitem{Nozieres1963}
Philippe \bgroup\fonteauteurs\bgroup Nozières\egroup\egroup{} :
\newblock {\em Le problème à $N$ corps\,:\,propriétés générales des gaz
  de fermions}.
\newblock Dunod, Paris, 1963.

\bibitem{Gradshteyn}
I.~S. \bgroup\fonteauteurs\bgroup Gradshteyn\egroup\egroup{} et I.~M.
  \bgroup\fonteauteurs\bgroup Ryzhik\egroup\egroup{} :
\newblock {\em {Tables of Integrals, Series, and Products}}.
\newblock {Academic Press}, San Diego, 1994.

\bibitem{Matera2010}
V.I. \bgroup\fonteauteurs\bgroup Abrosimov\egroup\egroup{}, D.M.
  \bgroup\fonteauteurs\bgroup Brink\egroup\egroup{},
  A.~\bgroup\fonteauteurs\bgroup Dellafiore\egroup\egroup{} et
  F.~\bgroup\fonteauteurs\bgroup Matera\egroup\egroup{} :
\newblock Self-consistency and search for collective effects in semiclassical
  pairing theory.
\newblock {\em Nuclear Physics A}, 864(1)\string:\penalty500\relax 38 -- 62,
  2011.

\bibitem{Zhang2016}
Boyang \bgroup\fonteauteurs\bgroup Liu\egroup\egroup{}, Hui
  \bgroup\fonteauteurs\bgroup Zhai\egroup\egroup{} et Shizhong
  \bgroup\fonteauteurs\bgroup Zhang\egroup\egroup{} :
\newblock {Evolution of the Higgs mode in a fermion superfluid with tunable
  interactions}.
\newblock {\em Phys. Rev. A}, 93\string:\penalty500\relax 033641, mars 2016.

\bibitem{CarusottoCastin2005}
Iacopo \bgroup\fonteauteurs\bgroup Carusotto\egroup\egroup{} et Yvan
  \bgroup\fonteauteurs\bgroup Castin\egroup\egroup{} :
\newblock {Atom Interferometric Detection of the Pairing Order Parameter in a
  Fermi Gas}.
\newblock {\em Phys. Rev. Lett.}, 94\string:\penalty500\relax 223202, juin
  2005.

\end{thebibliography}
\bibliographystyle{unsrtnat}

\end{document}


\title{Branche d'excitation collective du continuum dans les supraconducteurs BCS et les gaz de fermions superfluides}

\author{H. Kurkjian, S. Klimin et J. Tempere}
\affiliation{TQC, Universiteit Antwerpen, Universiteitsplein 1, B-2610 Anvers, Belgique}

\author{Y. Castin}
\affiliation{Laboratoire Kastler Brossel,
École Normale Supérieure, Université PSL, CNRS, Sorbonne Université, Collège de France, Paris, France}

\begin{abstract}
Nous montrons l'existence d'une branche d'excitation collective dans le continuum de brisure de paires des gaz de fermions superfluides et des supraconducteurs BCS. Nous prolongeons analytiquement l'équation de la RPA donnant l'énergie des modes collectifs à travers sa ligne de coupure associée au continuum, et obtenons l'intégralité
de la relation de dispersion complexe, y compris dans le régime de couplage fort. Pour $\Delta/\mu>1\vd210$ (très proche de la limite unitaire dans un gaz de fermions superfluide), où $\Delta$ est le paramètre d'ordre et $\mu$ le potentiel chimique, la partie réelle de la branche est intégralement située dans la bande interdite $[0,2\Delta]$.
Dans la limite des grandes longueurs d'onde, la branche varie quadratiquement avec le nombre d'onde, avec une masse effective complexe que nous calculons analytiquement. Ceci contredit le résultat de Littlewood et Varma mais confirme celui d'Andrianov et Popov.
\end{abstract}
\maketitle

\textit{Introduction --} Les systèmes dans lesquels une cohérence macroscopique s'établit entre paires de fermions présentent naturellement dans leur spectre d'excitation un continuum lié à la rupture de ces paires, dont l'énergie est supérieure à deux fois le paramètre d'ordre $\Delta$. C'est le cas notamment des supraconducteurs et des gaz froids d'atomes fermioniques de spin $1/2$. Le comportement collectif des gaz neutres aux énergies inférieures à $2\Delta$ est connu : il se caractérise par une branche d'excitation bosonique, de départ phononique \cite{Anderson1958}. La relation de dispersion a été calculée dans l'approximation de la phase aléatoire (RPA) \cite{Strinati1998,CKS2006} 
et l'existence de la branche confirmée expérimentalement \cite{Tachiki1997,Thomas2007,Vale2017}. 

À l'inverse, la question de l'existence d'un mode collectif \textit{dans} le continuum de brisure de paires, qui suscite un fort intérêt en raison d'une analogie souvent avancée avec le mode de Higgs en théorie des champs \cite{Varma2015}, n'a pas reçu de réponse définitive. L'enjeu est de comprendre si la réponse du système à une excitation est plate en fréquence ou si elle présente une structure non triviale telle qu'une résonance. Nous identifions deux lacunes majeures dans le traitement théorique \cite{Schmid1968,Orbach1981,Popov1976,Popov1987-III13,Varma1982} de ce mode : (i) il repose sur une hypothèse de symétrie particule-trou et néglige le couplage entre l'amplitude et la phase du paramètre d'ordre, ce qui le restreint au régime de couplage faible, (ii) il est limité aux grandes longueurs d'onde. Ces lacunes sont préjudiciables car elles engendrent des doutes sur l'existence même de ce mode \cite{Benfatto2016}, en particulier à vecteur d'onde nul \cite{Stringari2012}. 

Dans cet article, nous clarifions considérablement la description des modes collectifs du continuum. En prolongeant analytiquement la matrice des fluctuations gaussiennes du paramètre d'ordre, nous faisons apparaître un pôle dans le propagateur de $\Delta$ sous la ligne de coupure associée au continuum, pour un potentiel chimique $\mu>0$ et à vecteur d'onde non nul uniquement. Nous obtenons la relation de dispersion complète de ce mode sans négliger le couplage amplitude-phase. Ceci nous permet de traiter le régime de couplage fort ; en particulier nous voyons la partie réelle de la branche être intégralement inférieure à $2\Delta$ lorsque $\Delta>1\vd210 \mu$ (la branche reste néanmoins séparée de la bande interdite physique $[0,2\Delta]$ sur l'axe réel par une ligne de coupure). Dans la limite de couplage faible et des grandes longueurs d'onde, nous sommes en accord avec le résultat de Popov et Andrianov \cite{Popov1976}, et donc en désaccord flagrant avec la prédiction communément acceptée dans la littérature \cite{Varma1982}, en particulier pour le taux d'amortissement que nous trouvons de départ quadratique en le {nombre} d'onde, et non pas linéaire. Toutes nos prédictions sont fondées sur la RPA pour des interactions de contact. Cette théorie décrit qualitativement bien tant les gaz de fermions froids dans le raccordement CBE-BCS que les supraconducteurs BCS (les interactions de Coulomb sont sans effet sur les modes collectifs d'amplitude à des fréquences $O(\Delta/\hbar)$ \cite{Popov1976}) 
et est un préalable à toute description plus réaliste des interactions.

 {La branche que nous trouvons décrit le comportement collectif des paires suite à une excitation de leurs degrés de liberté internes ; sa pulsation n'est donc pas trivialement celle du bord du continuum $2\Delta/\hbar$, comme pour les « oscillations de Higgs » prédites et observées \cite{Kogan1973, Altshuler2006, Gurarie2009, Stringari2012, Foster2015, Klein1980, Shimano2013, Sacuto2014, Devereaux2015, Koehl2018} à vecteur d'onde nul.} 
{Elle est observable dans un gaz de fermions superfluide sous la forme d'un pic élargi dans la fonction de réponse du module du paramètre d'ordre.}

\textit{Fluctuations du paramètre d'ordre --}
Nous considérons un système homogène de fermions de masse $m$, de spin $\sigma=\uparrow/\downarrow$ et de potentiel chimique $\mu$, en interaction de contact. 
À température nulle $T=0$, les fluctuations du paramètre d'ordre $\Delta$ autour de sa valeur d'équilibre admettent des modes propres, qui sont les modes collectifs du système. En développant au second ordre en les fluctuations d'amplitude $\delta\lambda$ et de phase $\delta\theta$ on obtient l'action gaussienne \cite{Randeria1997,Randeria2008}
\begin{equation}
	\mathcal{S} = \mathcal{S}_0 + \int \dd \omega \int \dd^3 q 
		\begin{pmatrix}
			 -\ii\Delta\delta\theta_{-\qq} & \delta\lambda_{-\qq}
		\end{pmatrix} {{M}}(\omega, \bf{q})
		\begin{pmatrix}
			\ii\Delta\delta\theta_{\qq} \\ \delta\lambda_{\qq}
		\end{pmatrix}
\end{equation}
La matrice $M$ des fluctuations {est symétrique} et donne accès au propagateur du champ $\Delta$ par une simple inversion. L'équation donnant l'énergie $z_\qq$ des modes collectifs en fonction de leur vecteur d'onde $\qq$ s'écrit donc
\be
\mbox{det} M(z_\qq,\qq)=0 
\label{eq:RPA}
\ee
Le paramètre d'ordre $\Delta$ étant relié à la condensation de paires, les coefficients de sa matrice de fluctuations contiennent une intégrale sur le vecteur d'onde interne $\kk$ des paires, faisant intervenir $\xi_k=\hbar^2 k^2/2m-\mu$ et $E_k=\sqrt{\xi_k^2+\Delta^2}$ les relations de dispersion des fermions libres et des quasi-particules BCS respectivement, ainsi que l'énergie d'une paire d'excitations de vecteur d'onde total $\qq$, $E_{\kk\qq}=E_{\kk+\qq/2}+E_{\kk-\qq/2}$ :
\bea
\!\! M_{\pm\pm}(z,\qq) \!\! &=& \!\! \int \frac{\dd^3k}{2}\bbcro{\frac{(W_{\kk\qq}^\pm)^2}{z-E_{\kk\qq}}-\frac{(W_{\kk\qq}^\pm)^2}{z+E_{\kk\qq}} +\frac{1}{E_\kk}} \label{eq:Mpp}\\
\!\! M_{+-}(z,\qq)        \!\! &=& \!\! \int \frac{\dd^3k}{2}W_{\kk\qq}^+ W_{\kk\qq}^- \bbcro{\frac{1}{z-E_{\kk\qq}}+\frac{1}{z+E_{\kk\qq}} } \label{eq:Mpm}
\eea
où les indices $+$ et $-$ se rapportent respectivement aux fluctuations de phase et d'amplitude et nous introduisons la notation $(W_{\kk\qq}^\pm)^2=(E_{\kk+\qq/2}E_{\kk-\qq/2}+\xi_{\kk+\qq/2}\xi_{\kk-\qq/2}\pm\Delta^2)/(2E_{\kk+\qq/2}E_{\kk-\qq/2})$.\footnote{Avec $W_{\kk\qq}^+>0$ pour tout $\kk$ et $W_{\kk\qq}^->0$ si et seulement si $k^2>2m\mu/\hbar^2-q^2/4$.}  
Relevons que l'équation \eqref{eq:RPA} se retrouve à l'identique dans la RPA \cite{Anderson1958,artmicro,TheseHK}, par resommation de diagrammes \cite{CKS2006} ou par linéarisation des équations BCS dépendant du temps \cite{Annalen}. 

L'équation \eqref{eq:RPA} étant invariante par changement de $z$ en $-z$, nous pouvons nous restreindre à {$\textrm{Re}\, z\geq0$}. La matrice $M$ admet alors une ligne de coupure pour $z\in\mathcal{C}_\qq=\{E_{\kk\qq},\kk\in\mathbb{R}^3\}$, provenant des termes ayant $z-E_{\kk\qq}$ au dénominateur dans (\ref{eq:Mpp}--\ref{eq:Mpm}). Telle quelle, l'équation \eqref{eq:RPA} admet {au plus} une solution à $\qq$ fixé : {celle-ci} est réelle, d'énergie inférieure au continuum $\hbar\omega_{{\rm B},\qq}<\mbox{min}\, \mathcal{C}_\qq$ et correspond à la branche bosonique d'Anderson-Bogolioubov \cite{CKS2006}. Les modes collectifs que nous voulons caractériser se trouvent au contraire \textit{dans} le continuum, c'est-à-dire \textit{a priori} pour $\mbox{Re}\,z_\qq>\mbox{min}\, \mathcal{C}_\qq$. Comme dans le cas d'école d'un atome couplé au rayonnement électromagnétique \cite{Cohen}, la bonne façon d'obtenir une solution à \eqref{eq:RPA} en présence du continuum est de prolonger analytiquement la matrice $M$ à travers sa ligne de coupure \cite{Schmid1968}. Ceci est l'occasion pour nous de rappeler la  procédure de Nozières \cite{Nozieres1963} pour prolonger analytiquement une fonction de la forme
\be
f(z)=\int_{-\infty}^{+\infty} \dd \omega \frac{\rho(\omega)}{z-\omega},
\label{eq:fz}
\ee
analytique pour $\mbox{Im}\,z\neq0$ mais présentant une ligne de coupure sur l'axe réel, en tout point où la densité spectrale $\rho(\omega)$ est non nulle. 
La contribution non analytique à $M_{\sigma\sigma'}$, avec $\sigma,\sigma'=\pm$, se met naturellement sous la forme \eqref{eq:fz} avec la densité spectrale
\be
\rho_{\sigma\sigma'} (\omega,\qq)=\int \frac{\dd^3 k}{2} {W_{\kk\qq}^\sigma W_{\kk\qq}^{\sigma'}} \delta(\hbar\omega-E_{\kk\qq})
\label{eq:rho}
\ee
Le prolongement analytique de $f$ du demi-plan complexe supérieur au demi-plan inférieur, à travers un intervalle $]\omega_1,\omega_2[$ de la ligne de coupure où la fonction $\rho$ est analytique, s'écrit simplement
\be
f_\downarrow(z)=\begin{cases} f(z) \quad \mbox{si} \quad \mbox{Im}\, z>0 \\ 
f(z) - 2\ii\pi\rho(z) \quad \mbox{si} \quad \mbox{Im}\, z\leq0 \end{cases}
\label{eq:prescription}
\ee
où $z\mapsto\rho(z)$ est le prolongement analytique de $\rho$ pour $\mbox{Im}\, z\neq0$.
On s'en convainc sans mal 
en écrivant $\rho(\omega)=(\rho(\omega)-\rho(z))+\rho(z)$ dans \eqref{eq:fz} après l'ajout d'une coupure en énergie.

\begin{figure}
\begin{center}
\includegraphics[width=0.48\textwidth]{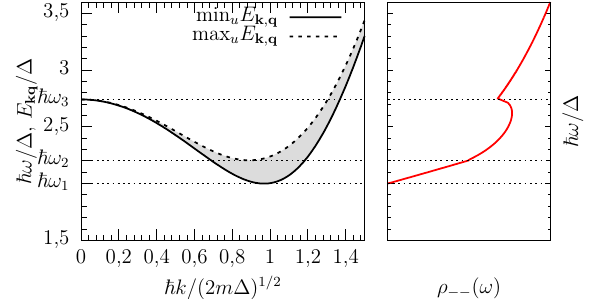}
\end{center}
\caption{\label{fig:energie} {En fonction de $k$, l'intervalle entre $\mbox{min}_u E_{\kk\qq}$ atteint en $u=0$ (trait plein à gauche) et $\mbox{max}_u E_{\kk\qq}$ atteint en $u=\pm1$ (tireté à gauche) définit une bande d'énergie (zone en grisé à gauche) sur laquelle la résonance $\hbar\omega=E_{\kk\qq}$ se produit pour au moins une valeur de $u=\cos(\kk,\qq)$ appartenant à $[-1,1]$.} À $\omega$ fixé, l'intervalle d'intégration sur $k$ dans \eqref{eq:rho} se lit horizontalement ; en fonction de $\omega$, sa structure subit 3 transitions en $\omega_1$, $\omega_2$ et $\omega_3$ (pointillés), qui se traduisent par des points anguleux dans la densité spectrale, ici l'exemple de $\rho_{--}$ (trait plein rouge à droite). {Sur cette figure, $\mu/\Delta=1$ et $\hbar q/\sqrt{2m\Delta}=0\vd5$.}}
\end{figure}

Pour effectuer le prolongement analytique de $M$, il nous faut donc étudier la fonction $\omega\mapsto\rho_{\sigma\sigma'}$ en premier lieu sur l'axe réel, 
en cherchant les singularités de la fonction et de ses dérivées. Pour cela, nous calculons l'intégrale sur $\kk$ dans \eqref{eq:rho} dans un repère sphérique d'axe $\qq$. L'intégrale sur l'angle azimutal disparaît par invariance par rotation et nous utilisons le Dirac pour effectuer l'intégrale sur l'angle polaire paramétré par $u=\kk\cdot\qq/kq$. L'intégrale restante sur $k$ est alors restreinte à un domaine représenté sur la figure \ref{fig:energie}, dont la forme dépend de $\omega$. 
Lorsque $\mu>0$  la branche d'excitation BCS possède son minimum en $k_0=\sqrt{2m\mu/\hbar^2}$ ; pour {$q>0$ suffisamment petit} \footnote{{Il existe en effet une valeur $q_0 < 2k_0$ telle que $k\mapsto\textrm{max}_u E_{\kk,\qq}$ ait son minimum en $k = 0$. Dans ce cas, deux des points
anguleux fusionnent : $\omega_3(q)=\omega_2(q)$. Puis, pour $q>2k_0$, les trois point anguleux sont confondus $\omega_3(q)=\omega_2(q)=\omega_1(q)>2\Delta/\hbar$.}}, la fonction $\omega\mapsto\rho_{\sigma\sigma'}$ possède alors trois points anguleux liés à un changement de configuration de ce {domaine}, ce qui découpe l'axe réel en quatre secteurs distincts (voir la figure \ref{fig:energie}). (i) Pour $\omega<\omega_1=2\Delta/\hbar$, la condition de résonance $\hbar\omega=E_{\kk\qq}$ n'est jamais vérifiée si bien que $\rho_{\sigma\sigma'}(\omega<\omega_1)=0$. Ce secteur est en dehors de la ligne de coupure $\mathcal{C}_\qq$.
 (ii) Pour $\omega_1<\omega<\omega_2$ la résonance est atteinte pour $k\in[k_1,k_2]$, avec
 \be
 \frac{\hbar k_{1,2}}{(2m)^{1/2}} = \sqrt{{\mu}-\frac{\hbar^2 q^2}{8m}\pm\frac{\sqrt{\hbar^2\omega^2-4\Delta^2}}{2}}
 \label{eq:k12}
\ee
(iii) Pour $\omega_2<\omega<\omega_3$, la résonance se produit pour $k\in[k_1,k_1']\cup[k_2',k_2]$ où $k_1$ et $k_2$ sont toujours donnés par \eqref{eq:k12} et $k_1'$ et $k_2'$ sont les solutions réelles positives du couple d'équations polynomiales
\be
\tilde{\omega}^4-4\tilde{\omega}^2P(\tilde{k},\tilde{q})+4R^2(\tilde{k},\tilde{q})=0 \quad  \mbox{et} \quad \tilde{\omega}^2-2P(\tilde{k},\tilde{q})>0
\label{eq:k12p}
\ee
avec les polynômes $P(x,y)=(x^2+y^2/4-\mu/\Delta)^2+x^2y^2+1$ et $R(x,y)=2xy(x^2+y^2/4-\mu/\Delta)$ et les notations $\tilde{k}^2=\hbar^2k^2/2m\Delta$, $\tilde{q}^2=\hbar^2q^2/2m\Delta$ et $\tilde{\omega}=\hbar\omega/\Delta$.
(iv) Pour $\omega>\omega_3$, les solutions $k_1$ et $k_1'$ ne sont plus réelles, l'intervalle d'intégration se réduit donc à $[k_2',k_2]$.

Une fois cette étude faite, l'extension des fonctions $\rho_{\sigma\sigma'}$ au plan complexe est immédiate, il suffit de remplacer $
\omega$ par $z$
dans les équations \eqref{eq:k12} et \eqref{eq:k12p} donnant les bornes d'intégration, $\rho_{\sigma\sigma'}(z)$ étant alors donnée par une intégrale curviligne entre des bornes complexes.  En appliquant la prescription de Nozières \eqref{eq:prescription}, on obtient ainsi un prolongement analytique de $M$, où les points anguleux $\omega_1$, $\omega_2$ et $\omega_3$ deviennent les points de branchement de trois lignes de coupure.

\textit{Étude numérique à $q$ quelconque --} Nous trouvons une solution $z_\qq=\hbar\omega_\qq-\ii\hbar{\Gamma_\qq}/{2}$ à l'équation \eqref{eq:RPA} dans le prolongement analytique à travers l'intervalle $[\omega_1,\omega_2]$, que nous identifions comme l'énergie du mode collectif que nous cherchons. 
Dans ce secteur, les fonctions spectrales s'expriment analytiquement en termes des intégrales elliptiques complètes de première et deuxième espèce\footnote{{En posant $\ch\Omega=\hbar\omega/2\Delta$ et en faisant $\hbar=2m=1$, on a $\rho_{++}(\omega)=\frac{\pi\Delta}{q}E(\ii\sh\Omega)$, $\rho_{--}(\omega)=\rho_{++}(\omega)-\frac{\pi\Delta}{q}K(\ii\sh\Omega)$, $\rho_{+-}(\omega)=0$.}}.
\begin{figure}[htb]
\begin{center}
\includegraphics[width=0.5\textwidth]{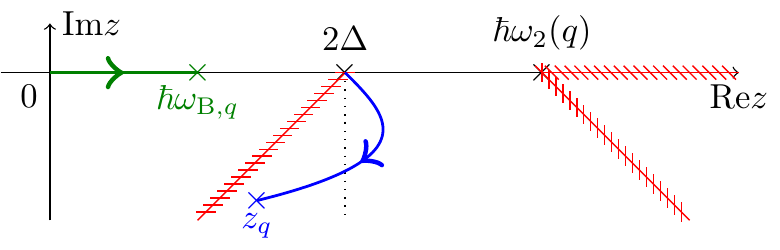}
\vspace{-1cm}
\end{center}
\caption{Trajectoires de la branche collective de brisure de paires 
(courbe bleue) et de la branche de Bogolioubov-Anderson
(courbe verte) dans le plan complexe, en fonction de $q$. La première de ces branches ne s'obtient qu'après prolongement analytique, d'où la ligne de coupure déformée (ligne rouge striée) dans le demi-plan inférieur. \label{fig:schema}}
\end{figure}

\begin{figure}
\begin{center}
\includegraphics[width=0.5\textwidth]{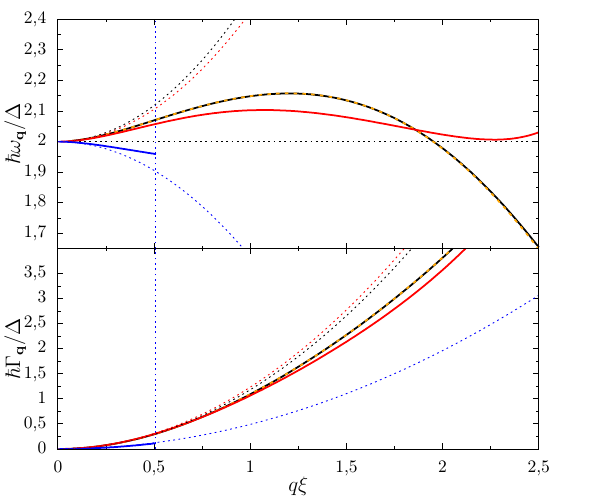}
\end{center}
\caption{\label{fig:disp} Pulsation (en haut) et taux d'amortissement (en bas) des modes collectifs du continuum en fonction de leur nombre d'onde $q$ en unité de la taille inverse des paires $1/\xi$, pour $\mu/\Delta=100$ ($1/k_{\rm F}a\simeq-3\vd0$, trait plein noir), $\mu/\Delta=5$ ($1/k_{\rm F}a\simeq-1\vd1$, trait plein rouge) {et $\mu/\Delta=0\vd1$ ($1/k_{\rm F}a\simeq0\vd5$, trait plein bleu)}. Le pointillé orange montre la relation dispersion obtenue pour $\mu/\Delta=100$ en négligeant le couplage module phase $M_{+-}$. Les courbes pointillées montrent le comportement quadratique en $q$ aux faibles $q$ obtenu analytiquement des équations \eqref{eq:reparam}-\eqref{eq:apres}.}
\end{figure}

La relation de dispersion $q\mapsto\omega_\qq$ est représentée sur la figure \ref{fig:disp} pour différentes forces d'appariement $\mu/\Delta=1/10$, $5$ et $100$ ($1/k_{\rm F}a\simeq0\vd5$, $-1\vd1$ et $-3\vd0$ dans un gaz de fermions froids \revision{de spin $1/2$, avec $k_{\rm F}$ le nombre d'onde de Fermi et $a$ la longueur de diffusion dans l'onde $s$}). 
De départ quadratique à partir de sa limite $2\Delta$ en $q=0$, la branche passe, en régime de couplage faible $\Delta\ll\mu$, par un maximum dont la hauteur en énergie est proportionnelle à $\Delta$ et dont la position est de l'ordre de l'inverse du rayon des paires $\xi\approx \hbar^2k_0/m\Delta$. Elle plonge ensuite sous $2\Delta$. Dans le régime de couplage fort $\Delta>\mu$, le domaine $[0,q_{\rm sup}]$ où l'énergie de la branche est supérieure à $2\Delta$ se réduit, jusqu'à sa disparition pour $\mu/\Delta\simeq0\vd8267$. Le taux d'amortissement $\Gamma_\qq$ est lui une fonction strictement croissante de $q$, de départ quadratique à partir de sa limite de valeur nulle en $q=0$. Ceci est en désaccord marqué avec la prédiction communément acceptée dans la littérature d'un taux d'amortissement linéaire en $q$ \cite{Varma1982}. 
{Le fait que la solution s'éloigne autant de la ligne de coupure initiale souligne la nature non perturbative du prolongement analytique que nous faisons : il n'existe pas de solution non perturbée sur l'axe réel dont $\textrm{Im}z_\qq$ pourrait se déduire par la règle d'or de Fermi.}

{La branche disparaît finalement en $q=2k_0$ (donc avant que la branche de Bogolioubov-Anderson ne rejoigne le continuum \cite{CKS2006}) lorsque l'intervalle $[\omega_1,\omega_2]$ à travers lequel passe notre prolongement analytique se réduit à un point.}
Enfin nous excluons l'existence d'une branche d'énergie supérieure à $2\sqrt{\Delta^2+\mu^2}$ dans le régime CBE où $\mu<0$ et où les trois singularités de $\rho_{\sigma\sigma'}$ se regroupent, $\omega_1=\omega_2=\omega_3=2\sqrt{(|\mu|+\hbar^2q^2/2m)^2+\Delta^2}/\hbar$.

\textit{Limite des grandes longueurs d'ondes --} Dans cette limite, nous obtenons plusieurs résultats analytiques qui corroborent notre étude numérique. Nous traitons à part le cas singulier $q=0$, où {la matrice $M(z,\qq=0)$ est exprimable} en termes des fonctions elliptiques complètes de première et troisième espèce, $K(k)$ et $\Pi(n,k)$ \cite{Gradshteyn}~:{
\bea
\mbox{th} s \tilde{M}_{++}(z,0) \!\!&=&\!\! \frac{\tilde{M}_{--}(z,0)}{\mbox{th} s} = - \pi(2 \eee^l)^{1/2} [F(s)-F(-s)] \notag \\
\tilde M_{+-}(z,0) \!\!&=&\!\! - \pi(2 \eee^l)^{1/2} [F(s) + F(-s)] \label{eq:Mpmq0}
\eea
avec les variables définies par $\mu/\Delta=\sh l$ et $z=2\Delta\ch s$, $s$ contraint par $\re s\geq 0$ et $-\pi/2\leq \im s \leq \pi/2$ et\footnote{{À l'aide des densités spectrales $\rho_{\sigma \sigma'}(\omega)$, on exprime $M_{\sigma\sigma'}(z,0)$ sous forme d'une intégrale sur $\omega$, en distinguant pour $\mu>0$ les contributions de $k<k_0$ et $k>k_0$. Le changement de variable $\hbar\omega = \Delta (x^2+1/x^2)$, $0<x<1$, fait alors disparaître la racine carrée $(\hbar^2\omega^2-4\Delta^2)^{1/2}=\Delta(1/x^2-x^2)$ et conduit à l'intégrale $\int\dd x R(x)/\sqrt{P(x)}$, où $P(x)$ est un polynôme de degré 4 et $R(x)$ une fraction rationnelle, ramenable à une combinaison linéaire d'intégrales elliptiques (cf. \S 8.110 de \cite{Gradshteyn}).}}
\be
F(s)\!\!=\!\!(\sh l +\sh s)[\Pi(\eee^{l+ s},\ii\eee^l)-\Pi(-\eee^{l- s},\ii\eee^l)]+K(\ii\eee^l)\ch s
\ee
L'équation \eqref{eq:RPA} s'écrit alors simplement $F( s) F(- s)=0$.}
La seule solution de cette équation, même après prolongement analytique\footnote{La ligne de coupure $[2\Delta,+\infty[$ (respectivement $[2\sqrt{\Delta^2+\mu^2},+\infty[$) en $z$ donne une ligne de coupure en $ s$ $[0,+\infty[$  pour $\mu>0$ (respectivement $[-l,+\infty[$ pour $\mu<0$). Ainsi, $F(s)-F(-s)$ admet la limite non nulle $\ii\pi(1+\sqrt{1+\Delta^2/\mu^2})^{-1/2}/\sqrt{2}$ quand $s\to0$ avec $\im s>0$.} de $F( s)$, est {$ s=\pm\ii\pi/2$} (soit $z=0$) : c'est {le} point de départ de la branche phononique d'Anderson-{Bogolioubov}. Ainsi, le point limite $\omega=2\Delta/\hbar$ en $q=0$ de la branche collective du continuum n'est ni solution de l'équation \eqref{eq:RPA} (comme l'avait compris la référence \cite{Matera2010}), ni un pôle des fonctions de réponse : $F( s)$ admet une limite finie et non nulle $F(0)$ lorsque $z\to 2\Delta$ ($s\to0$ avec $\mbox{Im}\, s>0$). Pour $q=0$, il n'y a donc pas de mode collectif de pulsation $2\Delta/\hbar$, mais rien de plus que le bord du continuum de brisure de paires. 
C'est pourquoi, les oscillations du paramètre d'ordre décrites par les références \cite{Kogan1973,Orbach1981,Altshuler2006,Gurarie2009,Stringari2012,Foster2015} ne sont pas sinusoïdales (comme $\cos(2 \Delta t/\hbar + \phi)$) mais amorties en loi de puissance ($\cos(2 \Delta t/\hbar + \phi)/t^\alpha$, {$\alpha>0$}). 

\begin{figure}
\begin{center}
\includegraphics[width=0.48\textwidth]{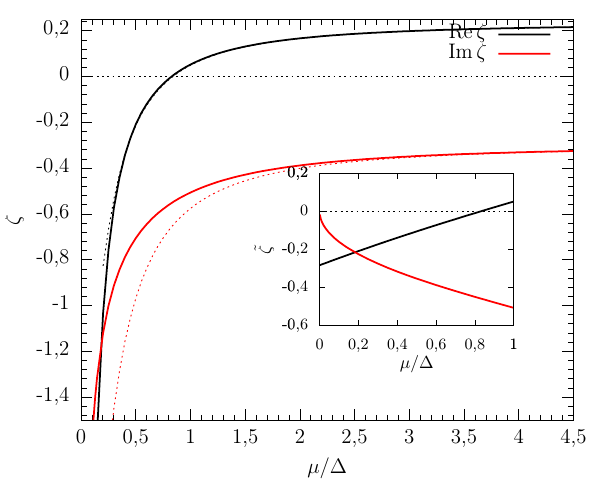}
\end{center}
\caption{\label{fig:z} La partie réelle (trait plein noir) et imaginaire (trait plein rouge) du coefficient adimensionné $\zeta$ de $q^2$ dans le développement de l'énergie $z_\qq$ du mode collectif de brisure de paires sont représentées en fonction de $\mu/\Delta$. Les courbes pointillées donnent le développement asymptotique dans la limite de couplage faible $\mu/\Delta\to+\infty$ : {$\zeta=\zeta_0-\frac{2 \zeta_0^2}{\zeta_0-1} \left(\frac{\Delta}{\pi\mu}\right)^2 \ln^2 \frac{\Delta}{8\mu \eee} + \ldots$ avec $\zeta_0=0\vd2369-0\vd2956\ii$}. L'encart montre le coefficient redimensionné $\tilde{\zeta}=\zeta \mu/\Delta=\zeta m/2m^*$ qui admet la limite finie réelle $\tilde{\zeta}_{\infty}=-16 K^2(\ii)/\pi^4\simeq-0\vd2823$ dans la limite de couplage fort $\mu/\Delta\to0^+$, {sa partie imaginaire tendant vers zéro comme $-12 K(\ii) (\mu/\Delta)^{1/2}/\pi^3$.}}
\end{figure}

Pour $q$ petit mais non nul, et $\mu>0$, le secteur de résonance entre $\hbar\omega_1=2\Delta$ et $\hbar\omega_2=2\Delta+\mu {\hbar^2q^2}/{2m\Delta}+O(q^4)$ sur la figure \ref{fig:energie} a une largeur en énergie $O(q^2)$, et $O(q)$ en le {nombre} d'onde $k$ autour du minimum $k_0=\sqrt{2m\mu/\hbar^2}$ de la branche BCS. Ceci suggère de poser
\be
z_\qq=2\Delta+\zeta\frac{\hbar^2q^2}{4m^*} + O(q^3) \quad \mbox{et} \quad k=k_0+Kq
\label{eq:reparam}
\ee
avec $m^*=m\Delta/2\mu$ la masse effective de la branche BCS au voisinage de son minimum. On concentre ainsi l'attention sur le domaine des vecteurs d'onde où le dénominateur dans (\ref{eq:Mpp},\ref{eq:Mpm}) est d'ordre $q^2$~:
\be
z-E_{\kk\qq} = z-2\Delta -\frac{\hbar^2 q^2}{m_*} (K^2 +u^2/4) + O(q^3)
\ee
En utilisant alors les développements des amplitudes du numérateur $W_{\kk\qq}^+\sim1$ et $W_{\kk\qq}^-\sim\hbar^2 k_0 q K/m\Delta$, puis en effectuant l'intégrale sur la variable angulaire $u$ avant celle sur $K$ (non sans avoir vérifié que le reparamétrage \eqref{eq:reparam} donne lieu lorsque $q\to0$ dans $M_{++}$ et $M_{--}$ à une intégrale absolument convergente sur $K$) on obtient sans mal les expressions analytiques pour $\mbox{Im}\, z>0$~:
\bea
\!\!\!\!\!\!\!\!\!\tilde{M}_{++}(z,\qq) \!\!\!&=&\!\!\! - \frac{\ii\pi^2 k_{\Delta}}{q}\asin \frac{1}{\sqrt{\zeta}} \label{eq:devMmm} +O(q) \label{eq:devMpp}\\
\!\!\!\!\!\!\!\!\!\tilde{M}_{--}(z,\qq) \!\!\!&=&\!\!\! -\frac{\ii\pi^2 \mu q}{2\Delta k_{\Delta}}\bbcro{\sqrt{\zeta-1}+\zeta\asin \frac{1}{\sqrt{\zeta}}} \!+\!O(q^3) \label{eq:devMmm} 
\eea
où nous introduisons $k_{\Delta}=\sqrt{2m\Delta/\hbar^2}$ et adimensionnons les intégrales {$\tilde{M}_{\sigma\sigma'}= {M}_{\sigma\sigma'}\Delta/k_{\Delta}^3$}. 
Puisque la divergence de $M_{++}$ d'ordre $1/q$ est compensée par l'annulation de $M_{--}$ linéaire en $q$, il suffit de prendre la limite finie et non nulle \eqref{eq:Mpmq0} de $M_{+-}$ en $q=0$ et $\hbar\omega=2\Delta$ (soit $ s=0$)~:
{
\be
{ {\tilde{M}_{+-}(2\Delta,0)}}=-2\sqrt{2}\pi \eee^{l/2} \mbox{Re}F(0) \label{eq:devMpm}
\ee}
Il n'y a plus qu'à insérer les expressions {(\ref{eq:devMpp},\ref{eq:devMmm},\ref{eq:devMpm})} dans l'équation de la RPA \eqref{eq:RPA} et à prolonger analytiquement le produit $M_{++}M_{--}$ à travers sa ligne de coupure $[0,1]$ en $\zeta$ (correspondant au segment $[\hbar\omega_1,\hbar\omega_2]$ en $z$) grâce aux substitutions $\asin 1/\sqrt{\zeta} \to \pi-\asin 1/\sqrt{\zeta}$ et $\sqrt{\zeta-1}\to -\sqrt{\zeta-1}$ pour obtenir l'équation explicite mais transcendante sur $\zeta$: 
{
\begin{multline}
\bbcro{\pi-\asin \frac{1}{\sqrt{\zeta}}} \bbcro{\bb{\pi-\asin \frac{1}{\sqrt{\zeta}}}\zeta- \sqrt{\zeta-1}} \\ + \frac{2}{\pi^4\mu} \bb{\frac{\hbar^2}{2m}}^{3} {M}_{+-}^2(2\Delta,0)=0
\label{eq:apres}
\end{multline}}
Le prolongement vaut pour tout le demi-plan inférieur, donc y compris pour $\mbox{Re}\, z<2\Delta$ ($\mbox{Re}\,\zeta<0$). L'unique solution de l'équation \eqref{eq:apres} reproduit fidèlement le coefficient de $q^2$ dans la figure \ref{fig:disp} ; elle est représentée sur la figure \ref{fig:z} en fonction de $\mu/\Delta$. La partie réelle change de signe en {$\mu/\Delta\simeq 0\vd8267$}, ce qui confirme l'existence d'une branche d'énergie inférieure à $2\Delta$ dans le régime de couplage fort. {Il ne faut pas surinterpréter l'existence de solutions de pulsation inférieure à $2\Delta$ : elles demeurent séparées du segment $[0,2\Delta]$ de l'axe réel par une ligne de coupure, si bien que leur effet ne se fait sentir au voisinage de l'axe réel qu'aux énergies supérieures à $2\Delta$.}

Pour mieux comprendre la disparition de la branche collective en $q=0$, 
nous calculons le résidu matriciel de
$M_\downarrow(z,\qq)^{-1}$ en $z_\qq$ et trouvons qu'il s'annule linéairement en $q$: il devient proportionnel au projecteur sur le sous-espace des fluctuations de module $\begin{pmatrix}0&0\\0&1\end{pmatrix}$ avec un coefficient
\be
Z_\qq\underset{q\to 0}{\sim} \frac{\ii \hbar^4 q}{2 m^2 \pi^2} \frac{\pi-\asin\frac{1}{\sqrt{\zeta}}}{(\pi-\asin\frac{1}{\sqrt{\zeta}})^2+\frac{(\pi-\asin\frac{1}{\sqrt{\zeta}})\zeta-\sqrt{\zeta-1}}{2\zeta\sqrt{\zeta-1}}}
\ee
Ceci s'obtient en utilisant $\frac{\dd}{\dd z}\propto q^{-2}\frac{\dd}{\dd\zeta}$ dans les équations (\ref{eq:devMpp},\ref{eq:devMmm}). {$Z_\qq$ mesure le poids spectral du mode collectif par rapport au fond du continuum ; son annulation en $q=0$ signifie que la fonction de réponse du système à $N$ corps ne peut plus s'interpréter en termes d'une quasiparticule sur un fond incohérent.}

Dans la limite de couplage faible $\mu/\Delta\to +\infty$, $M_{+-}$ tend vers zéro à cause de l'antisymétrie 
$k\leftrightarrow 2k_0-k$ par rapport à la surface de Fermi (valide tant que $(k-k_0)\xi=O(1)$), si bien que l'équation de la RPA se réduit à $M_{++} M _ {--} = 0$ pour $ q\xi ={O}(1) $, {et que l'équation \eqref{eq:apres} sur $\zeta$ se réduit à la partie dépendant de $\zeta$ de la première ligne.} Le mode collectif de brisure de paires correspond bien à un pur mode d'amplitude (une racine de $M_{\downarrow--}$)\footnote{{Bien que $M_{--}(2 \Delta,0)=0$, il n'y a pas de mode d'amplitude en $q=0$, même dans cette limite de couplage faible, car $1/M_{--}(z,0)$ n'est pas méromorphe.}}, l'annulation de $M_{++}$ donnant le mode de phase phononique dans 
cette limite. La relation de dispersion quadratique du mode du continuum
\be
z_\qq \stackrel[\mu/\Delta\to+\infty]{q \to 0}{\simeq} 2\Delta { + (0\vd2369-0\vd2956\ii) \frac{\hbar^2q^2}{4m^*} }
\ee
{contredit la référence \cite{Varma1982} (même $\textrm{Re}\,\zeta$ diffère de la valeur {$1/3$} de \cite{Varma1982}), mais confirme \cite{Popov1976}.}

\revision{Notre calcul montre les limites de l'analogie habituelle avec les modes de Higgs en théorie des champs : bien qu'il s'agisse aussi d'un mode d'amplitude dans la limite de couplage faible, et qu'il possède une bande interdite, le mode collectif, ici plongé dans un continuum, ne s'obtient qu'après traitement non perturbatif du couplage aux degrés de liberté fermioniques ; aucune chance donc de l'obtenir à partir d'une action effective de basse énergie comme il est suggéré parfois \cite{Varma2015,Zhang2016}.}

\begin{figure}
\begin{center}
\includegraphics[width=0.48\textwidth]{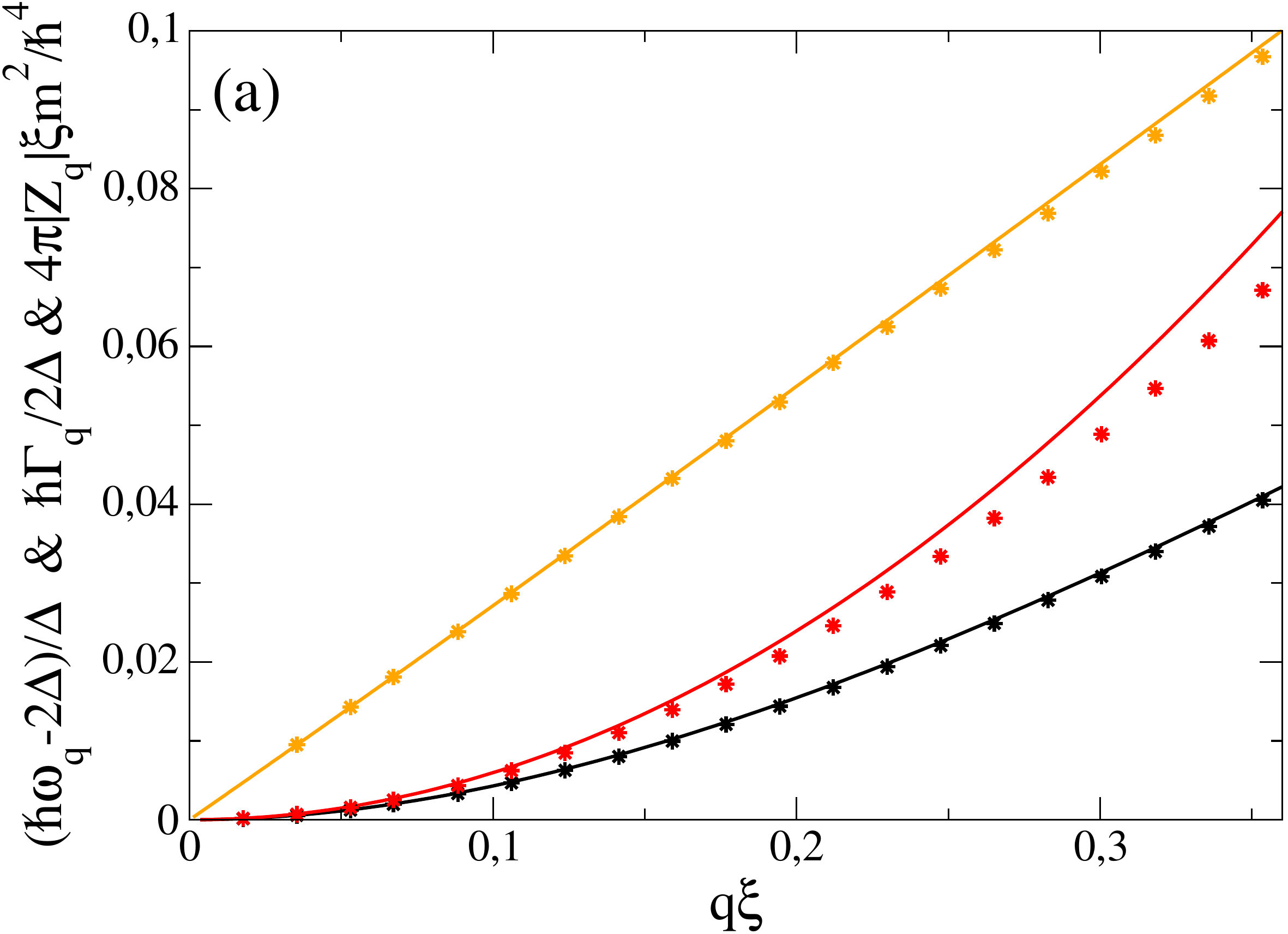}

\includegraphics[width=0.48\textwidth]{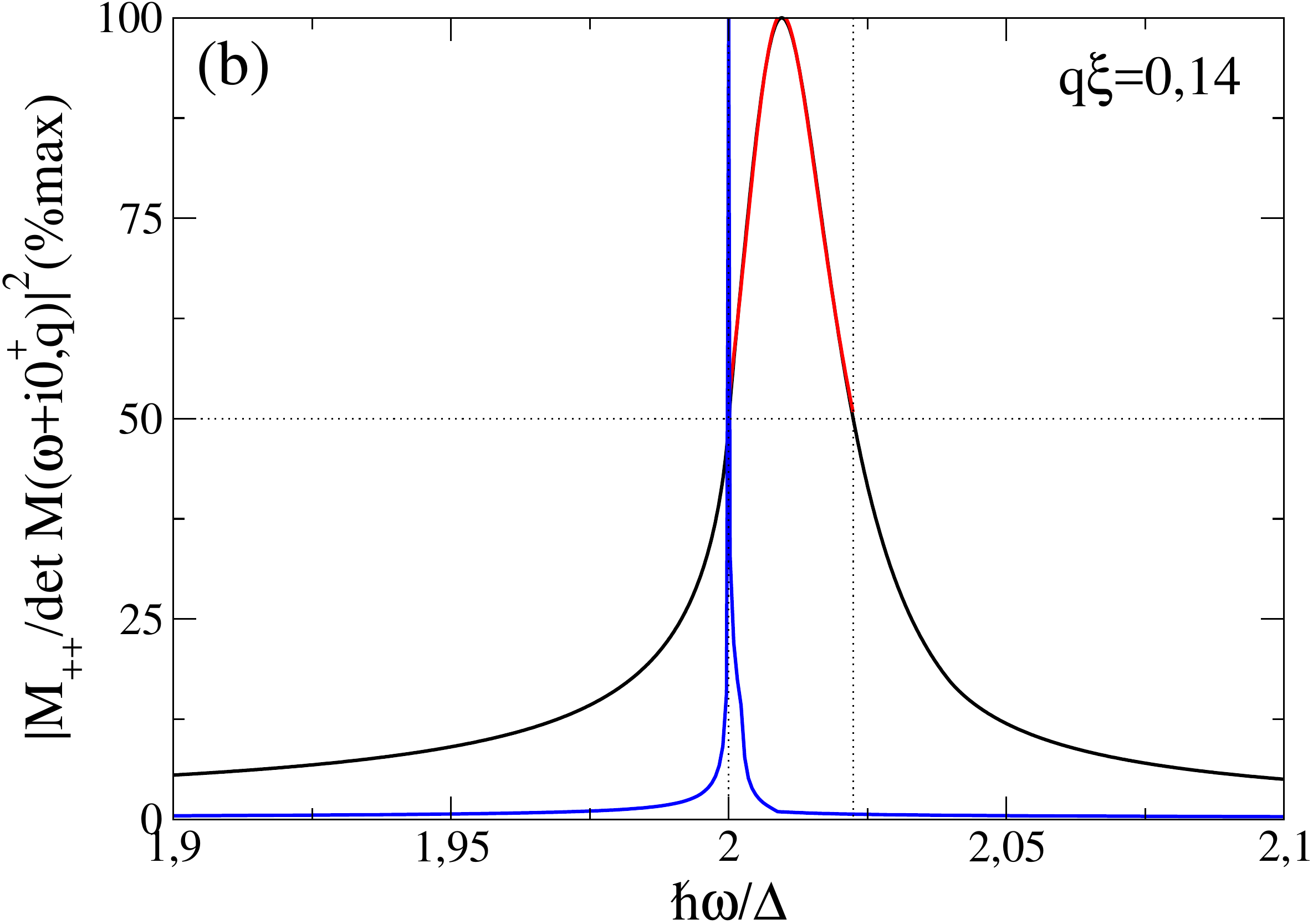}
\end{center}
\vspace{-0.5cm}
\caption{\label{fig:reponse} 
{$(a)$ En couplage faible ($\Delta/\mu\to0$), le déplacement de fréquence $\omega_\qq-2\Delta/\hbar$,  le taux d'amortissement $\Gamma_\qq$ et le résidu $Z_\qq$ du mode collectif de brisure de paires (respectivement courbe noire, rouge et orange) sont comparés aux valeurs extraites d'un ajustement de la fonction de réponse du module $1/|M_{--}(\omega+\ii0^+,\qq)|^2$ {(courbe noire du panneau $(b)$) par} la fonction $|C+Z_\qq^{\rm aj}/(\omega-\omega_\qq^{\rm aj}+\ii\Gamma_\qq^{\rm aj}/2)|^2$ {(courbe rouge de $(b)$)} décrivant une résonance simple sur un fond plat $C$. {Courbe bleue de $(b)$: fonction de réponse du module $|M_{++}/\textrm{det}M(\omega+\ii0^+,\qq)|^2$ en couplage fort ($\Delta/\mu=10$) où seul le pic pointu correspondant au bord du continuum  en $2\Delta/\hbar$ est visible.}}
}
\end{figure}

\textit{Observabilité dans les fonctions de réponse --} Aux faibles $q$, le mode collectif du continuum est faiblement amorti, condition favorable à son observation. En régime de couplage faible, il apparaît effectivement un pic lisse dans la fonction de réponse du module du paramètre d'ordre, comme le montre la figure \ref{fig:reponse}, et sa position, sa largeur et sa hauteur sont très bien prédites par la branche obtenue dans le prolongement analytique. En couplage fort (courbe bleue de la figure 5b), la résonance lisse a disparu et il ne reste qu'un pic pointu (admettant une tangente verticale), situé en $\omega=2\Delta/\hbar$ même pour $q\neq0$. Qualitativement, ceci indique que la fréquence $\omega_\qq$ du mode collectif est inférieure à $2\Delta/\hbar$ de sorte qu'il n'y a plus de résonance dans l'intervalle $[2\Delta/\hbar, \omega_2]$ que décrit notre prolongement analytique.

Contrairement à la fonction de réponse densité-densité souvent mesurée \cite{Vale2017}, la fonction de réponse du module du paramètre d'ordre ($ |M_{++}/\textrm{det}M(\omega+\ii0^+,\qq)|^2$, qui se simplifie en $1/|M_{--}(\omega+\ii0^+,\qq)|^2$ en couplage faible) est sensible au mode collectif de brisure de paires même en couplage faible. Dans les gaz d'atomes froids, le module du paramètre d'ordre peut être excité en modulant la force des interaction via une résonance de Feshbach, et mesuré par interférométrie résolue spatialement \cite{CarusottoCastin2005}.
Physiquement, la figure \ref{fig:reponse} montre que le système absorbe l'énergie venant d'une modulation de la force de l'appariement $|\Delta|$ à des fréquences $\omega>2\Delta/\hbar$ plus efficacement lorsque  $\omega$ est proche de $\omega_\qq$. La résonance est élargie car l'énergie absorbée peut être dissipée en brisant les paires en deux fermions désappariés de vecteurs d'onde $\qq/2\pm\kk$.

\textit{Conclusion --} Nous avons établi sur des bases théoriques solides l'existence d'une branche collective dans le continuum de brisure de paires d'un supraconducteur BCS ou d'un gaz de fermions superfluide, et nous avons caractérisé intégralement sa relation de dispersion et son taux d'amortissement dans l'approximation de la RPA, y compris dans le régime de couplage fort où le mode collectif est une superposition cohérente de phase et de module. Nous apportons ainsi une réponse claire et rigoureuse à un vieux problème de matière condensée. Notre branche apparaît nettement dans la fonction de réponse du module du paramètre d'ordre qui peut être mesurée dans un gaz d'atomes froids.

\bibliography{/Users/rox/Documents/biblio}
\bibliographystyle{unsrt-fr}